\documentclass[useAMS,usenatbib]{mn2e}
\usepackage{amssymb,amsmath,graphicx}
\usepackage{longtable}[1]
\usepackage{graphics}
\usepackage{threeparttable}
\usepackage{lscape}
\usepackage{color}
\usepackage{lscape}
\usepackage{rotating}

\def\msun{\hbox{M$_\odot$}}

\def\t4{\hbox{t$_{\rm 4}$}}

\def\myarcsec{\hbox{"}}
\def\mc{\hbox{M$_c$}}
\def\mt{\hbox{M$_T$}}

\def\relation{\hbox{SFR vs. M$_{V}^{\rm brigh}$}}

\def\cm3{\hbox{cm$^{-3}$}}

\def\ga{\hbox{$\Gamma$}}

\newcommand{\araa}{ARA\&A}
\newcommand{\apj}{ApJ}
\newcommand{\aj}{AJ}
\newcommand{\mnras}{MNRAS}
\newcommand{\aap}{A\&A}

\newcommand{\aaps}{A\&AS}

\newcommand{\apjs}{ApJS}

\newcommand{\ha}{H${\alpha}$}

\input psfig.sty
\input epsf.sty
\usepackage{graphicx}
\usepackage{epsfig}
\title[Cluster formation \& environment]
{Probing the role of the galactic environment in the formation of stellar clusters; using M\,83 as a test-bench}
\author[Adamo et al.] {A. Adamo$^{1}$\thanks{E-mail:
adamo@astro.su.se}, J. M. D. Kruijssen$^{2}$, N. Bastian$^{3}$, E. Silva-Villa$^{4}$, and J. Ryon$^{5}$\\
$^{1}$ Department of Astronomy, Oskar Klein Centre, Stockholm University, AlbaNova University Centre, SE-106 91 Stockholm, Sweden\\
$^{2}$ Max-Planck-Institut f\"ur Astrophysik, Karl-Schwarzschild-Strasse 1, 85748 Garching, Germany\\
$^{3}$ Astrophysics Research Institute, Liverpool John Moores University, 146 Brownlow Hill, Liverpool L3 5RF, UK\\
$^{4}$ Instituto de Fisica-FCEN, Universidad de Antioquia, Calle 70 No. 52-21, Medellin, Colombia \\
$^{5}$ Department of Astronomy, University of Wisconsin-Madison, 475 N. Charter St., Madison, WI, 53706, USA\\
}
\date{Accepted 2015 May 26. Received 2015 May 22; in original form 2015 March 04}
\pagerange{\pageref{firstpage}--\pageref{lastpage}}
\pubyear{2015}
\begin{document}
\maketitle
\label{firstpage}
\begin{abstract}
We present a study of the M\,83 cluster population, covering the disc of the galaxy between radii of 0.45 and 4.5 kpc. We aim to probe the properties of the cluster population as a function of distance from the galactic centre. We observe a net decline in cluster formation efficiency (\ga, i.e. the amount of star formation happening in bound clusters) from about 26\% in the inner region to 8\% in the outer part of the galaxy. The recovered \ga\, values within different regions of M 83 follow the same \ga\, versus star formation rate density relation observed for entire galaxies. We also probe the initial cluster mass function (ICMF) as a function of galactocentric distance. We observe a significant steepening of the ICMF in the outer regions (from $-1.90\pm0.11$ to $-2.70\pm0.14$) and for the whole galactic cluster population (slope of $-2.18\pm0.07$) of M\,83. We show that this change of slope reflects a more fundamental change of the 'truncation mass' at the high-mass end of the distribution. This can be modelled as a Schechter function of slope $-2$ with an exponential cut-off mass ($M_{\rm c}$) that decreases significantly from the inner to the outer regions (from 4.00 to $0.25\times 10^5$ \msun) while the galactic \mc\, is $\approx1.60\times10^5$ \msun. The trends in \ga\,  and ICMF are consistent with the observed radial decrease of the $\Sigma({\rm H}_2)$, hence in gas pressure. As gas pressure declines cluster formation becomes less efficient. We conclude that the host galaxy environment appears to regulate 1) the fraction of stars locked in clusters; 2) the upper mass limit of the ICMF, consistently described by a near-universal slope $-2$ truncated at the high-mass end.
\end{abstract}

\section{Introduction}
\label{sec:intro}
Young star clusters are potentially bright, long lived, tracers of star formation within their host galaxies. It is therefore of paramount importance to understand what governs the cluster formation process \citep[e.g.,][]{2014prpl.conf..291L} and how the host galactic environment may affect their formation and evolution \citep[e.g.,][]{aa..nb..2015Spr}.

Statistically we can describe the cluster formation as a stochastic process. Numerous works report of an initial cluster mass and luminosity function (ICMF and ICLF, respectively) being described with good approximation by a single power-law slope of index close to $-2$ \citep[e.g.,][]{2003A&A...397..473B, 2003MNRAS.343.1285D, 2007AJ....133.1067W, 2014AJ....147...78W, 2014ApJ...787...17C}. The stochastic nature of cluster formation combined with the power-law distributions of the cluster mass generate the so called size-of-sample effects. These effects are well described by the positive relations between the number of clusters \citep[][]{2000astro.ph.12546W} or the youngest brightest cluster in V band, M$_{V}^{\rm bright}$, versus the galactic star formation rate \citep[SFR;][]{2002AJ....124.1393L, 2008MNRAS.390..759B}. In other words, galaxies with higher SFR have more numerous cluster populations, therefore they are more likely to sample the ICMF at higher mass (luminosity) ranges. The size-of-sample effect also applies to the time-binning of the cluster population.  Massive clusters are statistically more likely to be older because the galaxy has had a longer period of time to sample the high mass bins of the ICMF \citep{2003AJ....126.1836H}. 

Increasing evidence, however, suggests that this is not the full picture. A steepening at the bright luminosity bins of the ICLF has widely been observed and reported \citep[e.g.,][]{1999AJ....118.1551W, 2002AJ....124.1393L, 2006A&A...446L...9G, 2012MNRAS.419.2606B, 2014AJ....147...78W}. This steepening could be explained by the presence of a truncation at the high mass end of the ICMF \citep{2006A&A...446L...9G}. Indeed, the presence of a truncation mass may explain why we do not observe very massive young star clusters close to $10^6$ \msun\, mass  in the Milky Way \citep{2006astro.ph..6625L}. However, it is difficult to establish the presence of a truncation due to low number statistics \citep{2006astro.ph..6625L}. Several works report an ICMF compatible with a Schechter function of slope $-2$, and an exponential cutoff at masses above a certain characteristic mass, \mc\, \citep[e.g.,][]{2009A&A...494..539L, 2012MNRAS.419.2606B, 2013AJ....145..137K}. Interestingly it has been observed by these authors that \mc\, may change as function of the host galactic environment. Such an environmental dependence would be expected theoretically due to the variation of the maximum mass scale for gravitational instability in galaxy discs (the "Toomre mass", Toomre 1964), which should lead to an increase of the maximum cluster mass with the gas pressure (Kruijssen 2014).

More evidence of the role played by the galactic environment on the properties of their cluster population has indirectly been suggested  by the so called T$_{\rm L}$(U) versus SFR relation \citep{2002AJ....124.1393L, 2011MNRAS.417.1904A}. T$_{\rm L}$(U) is the fraction of U band light locked in star clusters with respect to the total U band light of the galaxy. This fraction is observed to increase as function of the SFR of the galaxy, suggesting that the fraction of star formation happening in clusters (hereafter \ga\, or cluster formation efficiency, CFE) is increasing \citep{aa..nb..2015Spr}. Indeed direct evidence of a varying CFE as function of the average galactic star formation rate density ($\Sigma_{\rm SFR}$) has been observed in numerous galaxies \citep[e.g.,][among many others]{2010MNRAS.405..857G, 2011MNRAS.417.1904A, 2011AJ....142..129A, 2014AJ....148...33R}. 

In the model proposed by \citet{2012MNRAS.426.3008K} a variation in the CFE is directly linked to the star and cluster formation process itself. This model describes star and cluster formation continuously across the density spectrum of the hierarchically structured ISM, in which the highest star formation efficiencies (and thus bound stellar fractions) are reached in the highest density peaks.\footnote{In addition, tidal heating by encounters with giant molecular clouds (GMCs) imposes a minimum density for the formation of bound clusters, which in practice is only important in the very highest-density environments.} Integration of these local physics over the gas density spectrum then results in a CFE that increases with the gas surface density or pressure, as well as the SFR density (Kruijssen 2012). Given the density gradient observed in disc galaxies, the model thus predicts that the CFE decreases with the galactocentric radius. Indeed, using a small fraction of the whole cluster population of M\,83, \citet[][hereafter SV13]{2013MNRAS.436L..69S} report evidence of such a variation of the CFE as function of galactocentric distances. 

\begin{figure*}
\centering
\includegraphics[width=18cm]{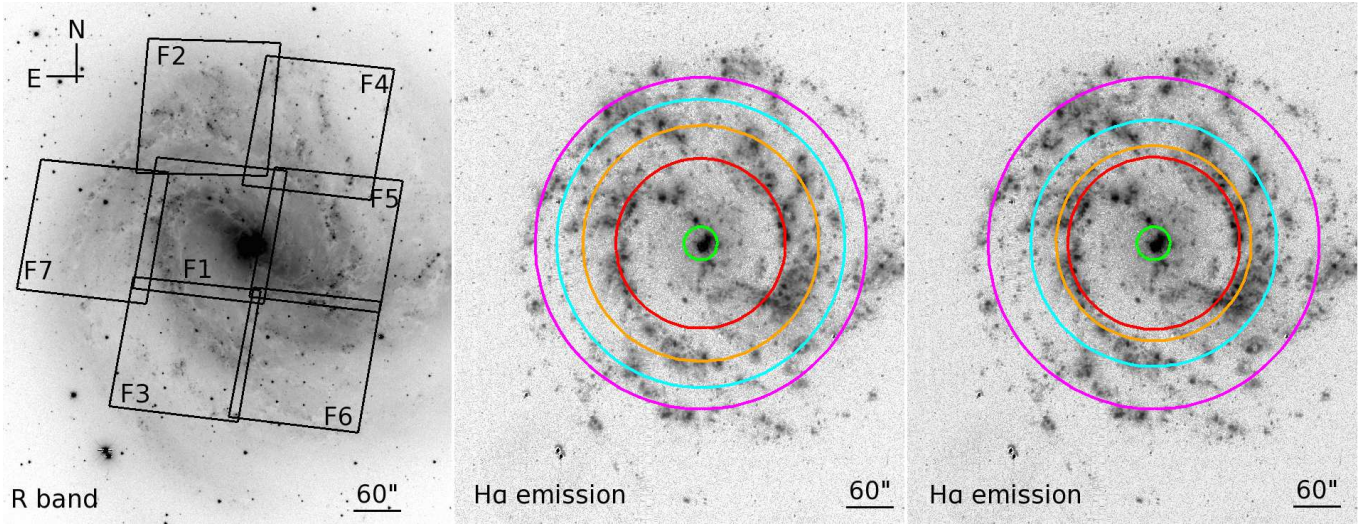}
\caption{Left: M\,83 image in the red visual band ($R$ band), the footprints of the seven HST pointings necessary to cover the bulk of the galactic body are overlaid. Centre and right: the H${\alpha}$ continuum subtracted image of M\,83 shows the regions where the current star formation is occurring in the galaxy. Annuli of equal area (centre) and containing equal number of clusters (right) are overplotted. The central region, contained inside the green circle, has been excluded from the analysis. Orientation and resolution of the images are indicated on each frame. Image credit at \citet{2009ApJ...703..517D}, taken from NASA/IPAC Extragalactic Database (NED).}
\label{fig:M83}
\end{figure*}

The aim of this article is to probe the imprints that the galactic environment leaves on the properties of its cluster population. The results will shed light on how the star cluster population is a natural outcome of the physics driving galactic star formation, and will also help to use star cluster populations as tracers of the SFH of their host galaxies. To achieve this goal we study the cluster population of the nearby ($\sim4.5$ Mpc,  corresponding to a distance modulus of 28.28 mag) spiral galaxy M\,83. This galaxy hosts a rich cluster population which has been widely analysed in recent years \citep{2010ApJ...719..966C, 2011A&A...529A..25S, 2011MNRAS.417L...6B, 2011ApJ...729...78W, 2012MNRAS.419.2606B, 2013MNRAS.436L..69S, 2014MNRAS.440L.116S, 2014ApJ...787...17C, 2015arXiv150203823H}. Several studies suggest that the current ongoing starburst at the centre of this galaxy \citep[e.g.,][]{2011ApJ...727..100W} was mostly likely produced by a recent minor merger \citep{2010MNRAS.408..797K}. Outside this region the most active star-forming regions in the galaxy are at the end of the bar (see Figure~\ref{fig:M83}). However, we notice that star formation is occurring over the entire spiral arm system and in spurs extended inside the inter-arm regions, as in M\,51 \citep{2011ApJ...727...88C}.

Only recently M\,83 has been fully observed with the Hubble Space Telescope (HST) with broad and narrow imaging bands covering from the $UV$ to the $NIR$ and several optical emission lines \citep{2014ApJ...788...55B}. Previous studies of the cluster population have therefore focused on the two pointings F1 and F2 (see Figure~\ref{fig:M83}). \citet{2010ApJ...719..966C, 2014ApJ...787...17C} report of a similar ICMF and cluster disruption rate in both inner (F1) and outer (F2) fields. They notice a steepening of the ICMF slope and a milder disruption rate  in F2, but they conclude that within the uncertainties the results support a cluster formation and evolution scenario independent of the galactic environment. With a similar cluster catalogue (comparison presented in both \citet{2012MNRAS.419.2606B, 2014ApJ...787...17C}), \citet[][hereafter B12]{2011MNRAS.417L...6B, 2012MNRAS.419.2606B} point out the differences in the photometric properties, ICMF and ICLF, disruption rate of the cluster populations in the two pointings. The differences cannot be accounted for by a recent, significant change in the SFH, so are therefore most likely linked to the cluster formation and disruption properties of the local galactic environment. Such a change in cluster properties is also supported by the change in the CFE as function of distance from the centre of the galaxy observed by SV13. 

The access to a full cluster catalogue for M\,83 now allows us to verify and put stronger constraints on previous findings. Using the whole M\,83 cluster catalogue, \citet[][hereafter SV14]{2014MNRAS.440L.116S} find a clear environmental dependency on the strength of the cluster disruption with higher cluster disruption in the inner regions of the galaxy, as already suggested by B12. In a recent publication, \citet{2015arXiv150203823H} has studied the \ha\, morphology of the very young star clusters in M\,83, reporting that clusters are already partially or fully exposed within a few Myr after their formation. Ryon et al (2015) has focused on the analysis of M\,83 cluster sizes, important for the dynamical evolution of these systems. In the present work we will expand the ICMF analysis published by B12 and the analysis of the CFE by SV13 to the whole cluster population within the M\,83 disk. All together, these works are presenting an unprecedentedly  detailed and complete study of the cluster population of a nearby galaxy.

The paper is organised as follows. Section 2 describes the photometric analysis and the properties of the final cluster catalogue. We also include a description of the ancillary data used in the analysis. In Section 3, we present our results. Section 4 contains a discussion of the results within the wider context of cluster and star formation in galaxies. A summary of the paper is presented in the Conclusions (Section 5).

\section{Description of the data}
\label{sec:obs}

\subsection{The cluster catalogue}

This analysis is based on the complete  M\,83 cluster catalogue recently published in SV14. The final cluster catalogue has been obtained from two multi-band HST imaging datasets, (GO 11360, PI: O'Connell, and GO 12513, PI: Blair). The two combined datasets cover within 7 pointings the face-on M\,83 disk up to a distance of 4.5 kpc (3.4\arcmin) from the centre of the galaxy (see Figure~\ref{fig:M83} and Table~\ref{tab1}). Our analysis is based on WFC3 imaging in 5 optical bands, F336W (hereafter $U$ band), F438W ($B$ band), F555W or F547M ($V$ band), F657N (narrow filter centred on \ha), and F814W (hereafter $I$ band). Cluster detection has been performed on the V band with optimised parameter inputs of the source extraction algorithm SExtractor \citep{1996A&AS..117..393B}. For each extracted source, aperture photometry has been performed in all the bands with an aperture radius of 5 px (0.2\myarcsec $\sim$ 4.4 pc) and a sky annulus located at 8 px (0.32\myarcsec) and 2 px (0.08\myarcsec) wide. Isolated compact clusters, visually selected, were used to estimate and add the amount of missed flux lost because of the small aperture size used. Moreover small differences in calibration between different pointings were corrected as explained in SV14. The final catalogue retains only sources with a concentration index larger than 1.25 mag and positive detections in 4 filters ($UBVI$). We consider detection in \ha\, as additional but not required constraint. The Vega zero points used to flux calibrate the photometry have been retrieved from the WFC3 instrument webpage\footnote{http://www.stsci.edu/hst/wfc3/phot\_zp\_lbn}. Milky Way foreground extinction correction has been applied \citep{2011ApJ...737..103S}. As described in SV14, the initial catalogue has been visually inspected and each source has been assigned a flag describing the morphology of the cluster. Class 1 objects are clusters with symmetric and compact light distribution. Class 1 objects are most likely bound clusters as tested in B12. Class 2 objects show asymmetries in the light distribution, multiple peaks, potentially contaminant neighbours. These objects closely resemble OB associations, therefore there is a fair chance of them being unbound \citep{2011MNRAS.410L...6G}. Class 3 objects are spurious detections (bad pixels mostly located at the end of the chip, background and foreground sources) and have been removed from the final catalogue. Any cluster candidate located within a radius of 0.45 kpc has also been excluded from the catalogue. We excluded this region because of the high luminosity gradient which causes a change in the completeness of the cluster detection and because of the different SFH that this region has experienced (B12). The number of cluster candidates detected in each field is listed in Table~\ref{tab1}.

\begin{table*}
  \caption{Description of the HST dataset used in this analysis. All the data have been taken with the WFC3.  The exposure time of each band is included in brackets. N$_{cl}$ is the final number of clusters (class 1 \& 2) in each field. We include within brackets how many of these clusters belong to class 1.}
\centering
\begin{threeparttable}
  \begin{tabular}{|c|c|c|c|}
  \hline
Field & PI & Filters & N$_{cl}\tnote{a}$\\
   \hline
F1 & O'Connell & F336W (1890 s), F438W (1880 s) & 1387 (672)\\
&&F555W (1230 s), F657N (1484 s) &\\
&&F814W (1213 s) & \\
   \hline
F2 & O'Connell & F336W (2560 s), F438W (1800 s) & 1098 (547)\\
&&F547M (1203 s), F657N (1484 s) &\\
&&F814W (1213 s) & \\
   \hline
F3 & Blair & F336W (2579 s), F438W (1799 s) & 1142 (595)\\
&&F547M (2682 s), F657N (1799 s) &\\
&&F814W (1379 s) & \\
   \hline
F4 & Blair & F336W (2589 s), F438W (1809 s) & 1243 (687)\\
&&F547M (2682 s), F657N (1809 s) &\\
&&F814W (1389 s) & \\
   \hline
F5 & Blair & F336W (2589 s), F438W (1809 s) & 1199 (676)\\
&&F547M (2682 s), F657N (1809 s) &\\
&&F814W (1389 s) & \\
   \hline
F6 & Blair & F336W (2579 s), F438W (1799 s) & 1414 (774)\\
&&F547M (2682 s), F657N (1799 s) &\\
&&F814W (1379 s) & \\
   \hline
F7 & Blair & F336W (2579 s), F438W (1799 s) & 859 (521)\\
&&F547M (2682 s), F657N (1799 s) &\\
&&F814W (1379 s) & \\
   \hline
\end{tabular}
 \begin{tablenotes}
 \item[a] The clusters located within 0.45 kpc from the centre have not been included in this final catalogue;
  \end{tablenotes}
 \end{threeparttable}
\label{tab1}
\end{table*}

\subsection{Physical cluster properties}

The observed spectral energy distribution (SED) of each object, reconstructed from the observed integrated fluxes in all the bands, has been compared to stellar evolutionary models including treatment of nebular gas emission and continuum \citep[\emph{Yggdrasil} models][]{2011ApJ...740...13Z}. The fitting algorithm is described in \citet{2010MNRAS.407..870A}. The model fluxes are reddened accordingly to the \citet{1989ApJ...345..245C} prescription. The algorithm simultaneously fits age and extinction of the cluster, while the mass is the normalisation factor to scale the models to the observed SEDs. Uncertainties on the derived ages, masses, and extinction are estimated accordingly to the recipe presented in \citet{2012MNRAS.426.1185A}. 

Although final cluster catalogues for F1 and F2 were available (B12) we decided to repeat the analysis for these two fields to make the dataset as homogeneous as possible. The comparison with the B12 catalogue shows some minor differences in the ages of the clusters between the latter catalogue and the new one as reported in SV14. The difference was produced by an erroneous extinction value systematically applied to the $I$ band which was a factor of 1.26 higher than the correct one. The cluster catalogue, published in this work, contains the correct and most updated cluster photometry and properties for M\,83. As already discussed in SV14 and in the present work, the main conclusions reached in B12 and SV13 are not affected by this error. 

\subsection{Estimates of the star formation rates}
\label{sfr}
In this work, we have estimated current (between 1 and 10 Myr) and average (between 10 and 50 Myr) star formation rates (cSFR and $\langle$SFR$\rangle$, respectively) in different regions of the galaxy as follows. The SFR is compared to the cluster formation rate (CFR) to derive \ga. Since clusters have been detected in optical wavelengths, we derive SFRs from optical based tracers. Thus, we use \ha\, as tracer of the cSFR and stellar counts to derive the recent SFH (up to 50 Myr). This choice may not take into account the hidden SFR better captured by an IR tracer, however, our SFR estimates will suffer similar extinction patterns as the clusters do, cancelling out the effects. 

To estimate the cSFR in different regions of the galaxy we used the publicly available \ha\, continuum subtracted image of M\,83 \citep[see][for details on the reduction process and calibration]{2009ApJ...703..517D}. We calibrated the total luminosity estimated from this frame (after masking residuals at the location of foreground stars) to the total \ha\, luminosity published by \citet{2008ApJS..178..247K}, giving us a correction factor that takes into account [N{\sc ii}] contamination and foreground galactic extinction. An average visual extinction of $A_V \sim 0.2$ mag has been applied to the total \ha\, flux. To convert \ha\, luminosity into SFR we use the relation by \citet{2012ARA&A..50..531K}.

To estimate the $\langle$SFR$\rangle$ we derive the recent SFH using colour-magnitude diagrams of the resolved stellar population. The technique and method are described in \citet{2012A&A...537A.145S} and SV13. Stellar photometry and analysis will be discussed in a forth-coming paper (Silva-Villa et al in prep.). For consistency with the cluster and \ha\, analyses, we corrected the final stellar photometry by an average visual extinction of 0.2 mag. Because of different detection limits between the arm and inter-arm regions we applied a conservative approach. Completeness tests show that our SFH is robust between 10 and 50 Myr, above and below which the significancy of our results drops. The $\langle$SFR$\rangle$ is thus derived averaging the SFH over 10 to 50 Myr.

\subsection{Estimates of the gas surface density}
\label{h2}

To estimate the gas surface density, $\Sigma({\rm H}_2)$, we use the velocity-integrated CO(1-0) intensity map published by \citet{2004A&A...413..505L}, to which we refer for a description of the dataset  and reduction steps. CO intensity was derived within each bin used in our analysis. To convert the CO intensity to a molecular gas mass, we adopt a conversion factor of $X_{\rm CO}=2.3\times10^{20}~{\rm K}~{\rm km}~{\rm s}^{-1}~{\rm cm}^2$.\footnote{We adopt this value for consistency with \citet{2004A&A...413..505L}. This choice is consistent with the Galactic conversion factor of $X_{\rm CO}=2.0\times10^{20}~{\rm K}~{\rm km}~{\rm s}^{-1}~{\rm cm}^2$ with an uncertainty of about 30\% \citep{2013ARA&A..51..207B}.} The derived values are in good agreement with averaged $\Sigma({\rm H}_2)$ reported in their work \citep{2004A&A...413..505L}.

\section{Results}
\label{sec:results}

\subsection{Analysis approach}

\begin{figure}
\centering
\includegraphics[width=8.5cm]{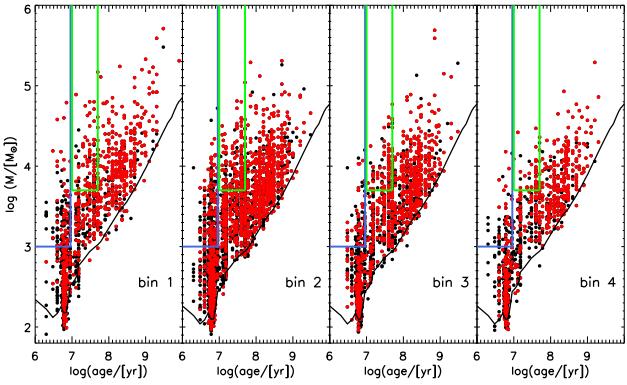}
\caption{Cluster age versus mass diagram of each bin of equal area. The black dashed lines show the detection limits in the $V$ band ($M_V\leq-5.5$ mag in bin 1 and $M_V\leq-5.3$ mag in the other 3 bins), the filter used to extract the source catalogue (see main text). Black dots are cluster candidates of class 2, red dots clusters belonging to class 1. The blue and green delimited areas show the clusters used to derive the current and averaged CFR. The limit on the cluster mass is well above the detection limits and has been imposed to mitigate the effect of stochasticity of the IMF.}
\label{fig:cl_pop}
\end{figure}

To understand how the galactic environment influences star and cluster formation we analyse and compare the cluster properties in different regions of M\,83. The galactic region contained between the inner radius of 0.45 kpc and outer radius of 4.5 kpc has been divided in four bins. As shown  in the central and right panels of Figure~\ref{fig:M83}, we choose two binning approaches, equal surface area bins and bins containing the same number of objects (the latter approach already applied  in SV13). For each bin we derive the following quantities (see Table~\ref{tab2}): cSFR (1 - 10 Myr); $\langle$SFR$\rangle$ (10 - 50 Myr); \ga(1 - 10 Myr) and \ga(10 - 50 Myr); predicted CFE (\ga$^{\rm th}$); gas surface density, $\Sigma({\rm H}_2)$; luminosity of the youngest most massive cluster, M$_{V}^{\rm bright}$; mass of the most massive cluster (in the age range 3 to 100 Myr), M$_{max}$; characteristic mass, \mc; and slope of the mass function if a single power-law is used, $\alpha$. In Figure~\ref{fig:cl_pop}, we show the age-mass diagram of the cluster population contained in each bin of equal area. We include both class 1 (red dots) and 2 (black dots) objects. We do not see any significant difference within the two classes and each bin has similar fractions of class1 versus class 2 objects. The blue and green boxes show the cluster population used to derived the CFE as described below. The age distributions of Figure~\ref{fig:cl_pop} clearly show the size-of-sample effect, i.e. the increasing M$_{max}$ as a function of age (and also of decreasing M$_{max}$ as a function of galactocenric distance).  

The CFE is defined as the ratio between the CFR and the SFR within the same age interval. We derive the CFR in the same interval used for the SFR (1-10 and 10-50 Myr). The CFR is the ratio between the total stellar mass in clusters and the corresponding age range. The current CFR is estimated from clusters of age between 1--10 Myr and cluster mass above a thousand \msun. This quantity is compared to the cSFR estimated within the same bin to derive \ga(1-10 Myr).  Similarly, the average CFR is estimated for clusters with age between 10 and 50 Myr and mass above $5000$ \msun\ and directly compared to the $\langle$SFR$\rangle$ thus deriving \ga (10-50 Myr).  To derive the total CFR we assume an underlying ICMF power-law of slope $?2$, with lower and upper mass limits of 100 \msun\, and two times the maximum observed cluster mass  in the region under investigation, respectively. The total CFR will be the result of the observed stellar mass in clusters more massive than the observational imposed limit and the missing fraction.

The two mass limits applied at different age ranges to estimate the CFR are well within the detection limits of our sample. They are chosen to be substantially higher than the detection limit to avoid completeness issues. However, these mass limits may cause possible misclassifications due to stochastic effects in the stellar initial mass function (IMF) as pointed out by several studies \citep[e.g.,][]{2010ApJ...724..296P, 2010A&A...521A..22F}. In a upcoming work, our deterministic approach to derive cluster properties is compared to the predictions obtained by SLUG (Krumholz et al in prep). SLUG is a bayesian fitting code that uses stochastic IMF models to derive cluster properties \citep{2015arXiv150205408K}.  Krumholz et al (in prep) find significant misclassification of the cluster mass below 1000 \msun, i.e., the deterministic approach used in this work significantly underestimates cluster masses below a 1000 \msun. 

We estimate the fraction of missing stellar mass in clusters with mass below the two thresholds, assuming a power-law slope distribution of $-2$ and a lower mass limit of 100 \msun\ \citep{2014prpl.conf..291L}. Only class 1 clusters are used to derive CFRs. Due to this conservative approach and the possible effects of stochasticity, our CFE determinations can be considered lower limits to the real values.  Monte Carlo simulations of cluster populations are used to estimate the uncertainties associated with the CFR (and hence \ga) values. We take into account a 0.1 dex error in both derived age and mass of each cluster (uncertainties from the SED fitting technique). We also include the poissonian likelihood of recovering the observed CFR. This uncertainty accounts for the observed number of clusters within the considered mass and age range as randomly drawn from a power-law cluster mass function with upper mass limit equal to two times the observed maximum mass value in each bin.

In Table~\ref{tab2} one can see that the SFR and \ga\, derived with different methods agree within 3 sigma in each bin, confirming that the SFR, as well as the CFE, has not changed significantly in the last 50 Myr (SV14). A different binning of the data does not produce any major impact in the derived quantities. As already discussed in SV13, bins containing the same number of clusters should mitigate the size-of-sample effect. In the M\,83 case, differences between equal surface area and equal number of cluster bins are mostly restricted to the regions containing a very active site of star-formation for the M\,83 disk, i.e. the edges of the bar (see the central and right panels of Figure~\ref{fig:M83}). Therefore, even if local variations between the two binning approaches are observed in the central bins covering the end of the bar region, the radial trends discussed in the next sections are preserved. 

\begin{table*}
  \caption{Galactic and cluster population properties derived for the whole population, in annuli of equal area, or containing equal number of clusters. Only class 1 clusters have been used in this analysis. The typical uncertainty on $\Gamma^{\rm theory}$ is a factor of two (see Section 3.2). The M$_{V}^{\rm bright}$, M$_{max}$, \mc, $\alpha$ have been derived for class1 clusters younger than 100 Myr and more massive than 5000 \msun.}
\centering
\begin{threeparttable}
  \begin{tabular}{|c|c|c|c|c|c|c|c|c|c|c|}
  \hline
annulus & cSFR & $\langle$SFR$\rangle$ & \ga$^{1-10 Myr}$&\ga$^{10-50 Myr}$&\ga$^{\rm th}$&$\Sigma({\rm H}_2)$& M$_{V}^{\rm bright}$& M$_{max}$&\mc&$\alpha$\\
 kpc&M$_\odot$yr$^{-1}$&M$_\odot$yr$^{-1}$& \%&\%&\%&M$_\odot$pc$^{-2}$& mag & $10^5$\msun&$10^5$\msun& \\
\hline
   \hline
   &&&&galactic&&&&&\\
   \hline
   0.45--4.50\tnote{a}& 0.82$\pm$0.8& 1.20$\pm$0.12& 18.2$\pm$3.0 & 15.0$\pm$1.9& 19.0\tnote{b} &47.7&  -11.60&  1.95$^{+      0.45}_{-      0.29}$&1.60$\pm$0.30&-2.18$\pm$0.07\\
   \hline
   \hline
         &&&&equal area&&&&&\\
   \hline
   0.45--2.30 &0.25$\pm$0.02&0.21$\pm$0.02&      19.3$\pm$4.0&      26.5$\pm$4.0&      28.6&      62.4&      -11.60& 1.48$^{+0.22}_{-0.19}$&       4.00$\pm$0.80&-1.90$\pm$0.11\\
   2.30--3.20 & 0.31$\pm$0.03&0.43$\pm$0.04&      26.0$\pm$3.5&      19.2$\pm$2.6&      20.4&      34.9&     -11.47&       1.95$^{+      0.45}_{-      0.29}$&      1.00$\pm$0.20&    -2.20$\pm$0.11 \\
   3.20--3.90 &0.15$\pm$0.02 & 0.34$\pm$0.03&      13.1$\pm$2.5&      9.8$\pm$1.6&      13.5&      19.4&     -10.68&      0.59$^{+      0.19}_{-      0.14}$&      0.55$\pm$0.11&   -2.20$\pm$0.12 \\
   3.90--4.50 &0.11$\pm$0.01 & 0.22$\pm$0.02&      13.7$\pm$3.1&      8.0$\pm$1.5&      12.6&      17.7&     -10.55&      0.29$^{+     0.04}_{-     0.02}$&      0.25$\pm$0.05&     -2.70$\pm$0.14 \\
   \hline
   \hline         
   &&&&equal number&&&&&\\
   \hline

 0.45--2.30 &  0.26$\pm$0.03 &      0.25$\pm$0.02 &     18.9$\pm$4.4 &       23.2$\pm$4.2 &      28.6 &       62.4 &    -11.60  & 1.48$^{+ 0.22}_{- 0.19}$ &       50.00$\pm$ 10.00  &     -1.90$\pm$0.11 \\
2.30--2.60 & 0.12$\pm$0.01 &      0.25$\pm$0.02 &       34.0$\pm$5.1 &       20.8$\pm$4.0 &       26.3 &       52.6 &      -11.47 &        1.95$^{+      0.45}_{-      0.29}$ &        1.50$\pm$0.30 &  -2.05$\pm$0.12 \\
2.60-3.30 & 0.21$\pm$0.02 &      0.37$\pm$0.04 &       19.8$\pm$3.6 &       10.8$\pm$2.1 &       17.4 &       27.6 &      -10.27 &       0.59$^{+      0.19}_{-      0.14}$ &       0.60$\pm$0.12 &  -2.35$\pm$ 0.13\\
3.30--4.50 & 0.23$\pm$0.02 &      0.33$\pm$0.03 &       14.0$\pm$2.6 &       12.6$\pm$2.2 &       13.0 &       18.3 &      -10.55 &       0.39$^{+     0.04}_{-     0.05}$ &       0.30$\pm$0.06 &    -2.50$\pm$0.13 \\
   \hline
   \hline
\end{tabular}
\begin{tablenotes}
\item[a] Properties of the galaxy and cluster population integrated within 0.45 and 4.5 kpc.
\item[b] Value from \citet{2012MNRAS.426.3008K}.
\end{tablenotes}
 \end{threeparttable}
\label{tab2}
\end{table*}

\subsection{On the fraction of star formation happening in clusters as function of distance within the galaxy}

\begin{figure}
\centering
\includegraphics[width=9cm]{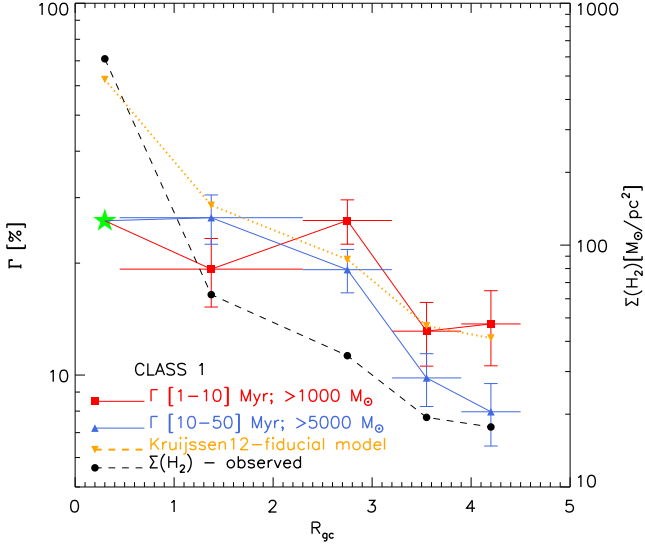}
\caption{Cluster formation efficiency as function of galactocentric distance. \ga\, has been estimated for two age ranges and different mass limits as indicated in the inset. Only clusters classified as class 1 have been used for the analysis. Therefore, the CFEs can be considered lower limits to the real fractions. \ha\, emission is the tracer used to estimate the SFR between 1 and 10 Myr, while direct stellar counts have been used to derive the SFR in the age range 10-50 Myr. The horizontal bars show the width of each annulus. The area within each annulus is the same. The green star shows the \ga\, value for the starburst region confined in the centre of the galaxy and reported by \citet{2010MNRAS.405..857G}. The right-hand y-axis shows the molecular gas surface density ($\Sigma({\rm H}_2)$). See text for more details.}
\label{fig:gamma_rad}
\end{figure}

In Figure~\ref{fig:gamma_rad} we show how \ga\, changes with galactocentric distances for bins of equal area. The decreasing trend observed here confirms the finding reported by SV13, but for a substantially larger cluster sample. The two \ga\, values (for each age range) follow each other quite closely. \ga\, decreases  inside out from about 26\% to 8\%. On different scales (right y-axis), we overplot  the gas surface density, $\Sigma({\rm H}_2)$ estimated within the same regions. $\Sigma({\rm H}_2)$ shows a similar decline (black dashed line) suggesting a link between the CFE and the amount of molecular gas available for star formation. The plot includes also the predictions for \ga\, obtained with the fiducial model by \citet{2012MNRAS.426.3008K}. The model predicts a higher absolute value than the observed one in the centre but agrees quite well with the values we find in the disc, within the uncertainties. We notice that the inner point (green star symbol), taken from \citet{2010MNRAS.405..857G}, is likely to be underestimated due to incompleteness caused by the strong differential extinction in this region. In the disc (between 0.45 and 4.5 kpc), the observed {\it relative} decline of \ga\, is closely reproduced by the model. 


The good agreement between the predicted and observed gradients of $\Gamma$ as a function of galactocentric distance suggests that the Kruijssen cluster formation model can be used to understand the cluster formation process. The model predicts that the formation of bound stellar clusters takes place in the highest-density peaks of a hierarchically structured ISM. At these high densities, the gas goes through a large number of free-fall times on a short time-scale and therefore reaches a high star formation efficiency. As a result, the expulsion of any residual gas by feedback does not unbind the stellar distribution \citep{kruijssen12a} and a bound cluster survives. At lower densities, the low star formation efficiencies yield unbound associations.\footnote{An additional included effect at high gas surface densities ($\Sigma>10^3~{\rm M}_\odot~{\rm pc}^{-2}$) is that tidal perturbations by dense giant molecular clouds unbind lower-density clusters that otherwise would have survived \citep[cf.][]{kruijssen11}.} Putting these local considerations in a galactic context, the model predicts that clusters form most efficiently at high gas pressures (and hence gas surface densities), because these conditions lead to higher density peaks and thus favour bound cluster formation. In summary, the model thus predicts that the CFE is set by galactic-scale properties such as the gas surface density and, therefore, the star formation rate density $\Sigma_{\rm SFR}$ through the Schmidt-Kennicutt relation \citep{2012ARA&A..50..531K}.

The two panels in Figure 4 support the above interpretation. We show the CFE as a function of $\Sigma_{\rm SFR}$ for our spatially resolved data points in M~83 as well as for a compilation of literature measurements of the CFE in entire galaxies. Both the entire galaxies and the different regions within galaxies follow the same trend of increasing CFE with the SFR density. We include the CFEs estimated for the two types of M~83 binning (equal area and equal cluster number bins). We also estimate the CFE in each of the 7 fields shown in Figure 1. We show the M~83 CFEs for both age ranges (left and right panel). Overall, the agreement with Kruijssen's fiducial model (black dotted line) is very encouraging and adds to the evidence supporting a tight link between galactic environment and stellar cluster formation.

\begin{figure*}
\centering
\includegraphics[width=7.5cm]{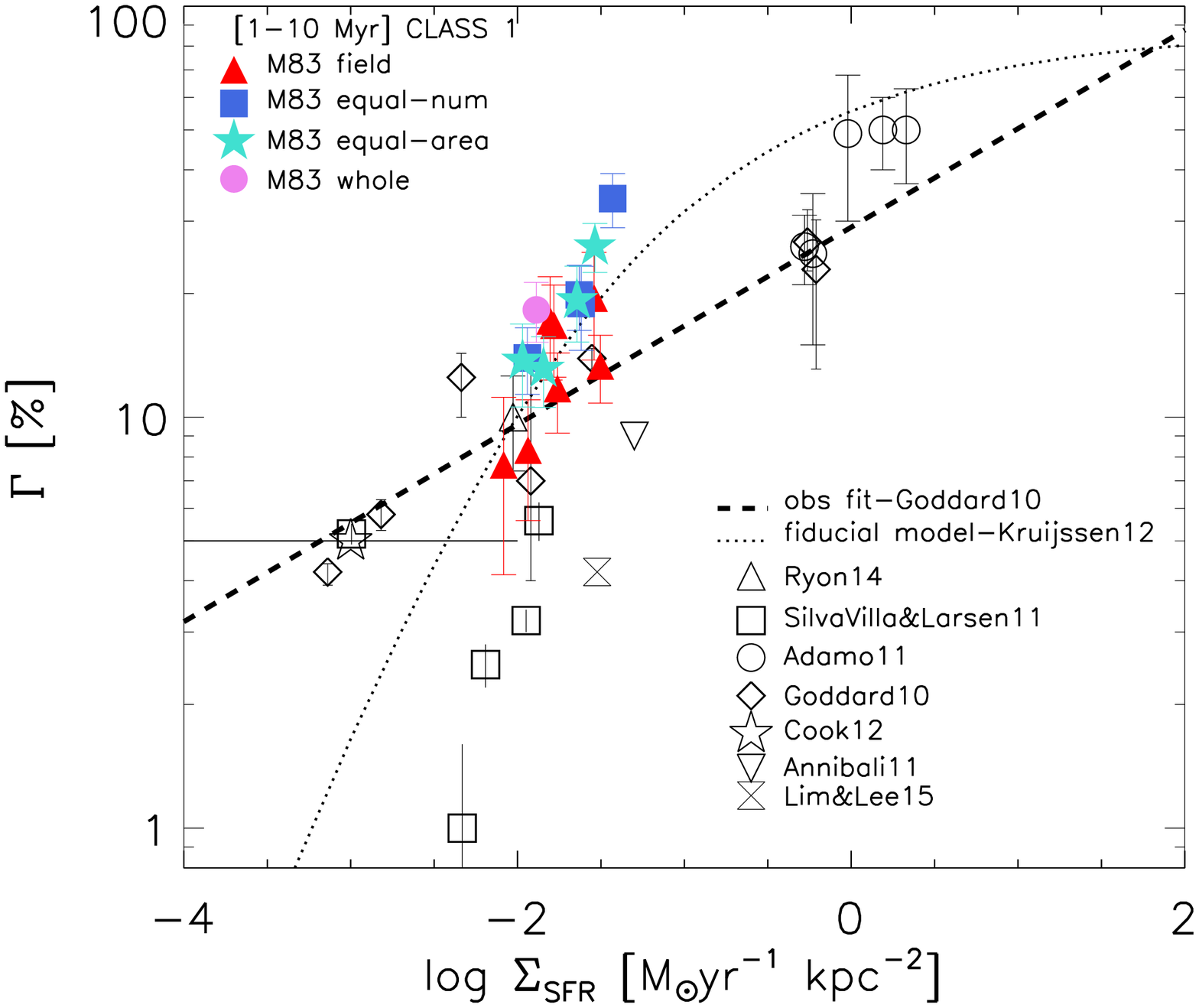}
\includegraphics[width=7.5cm]{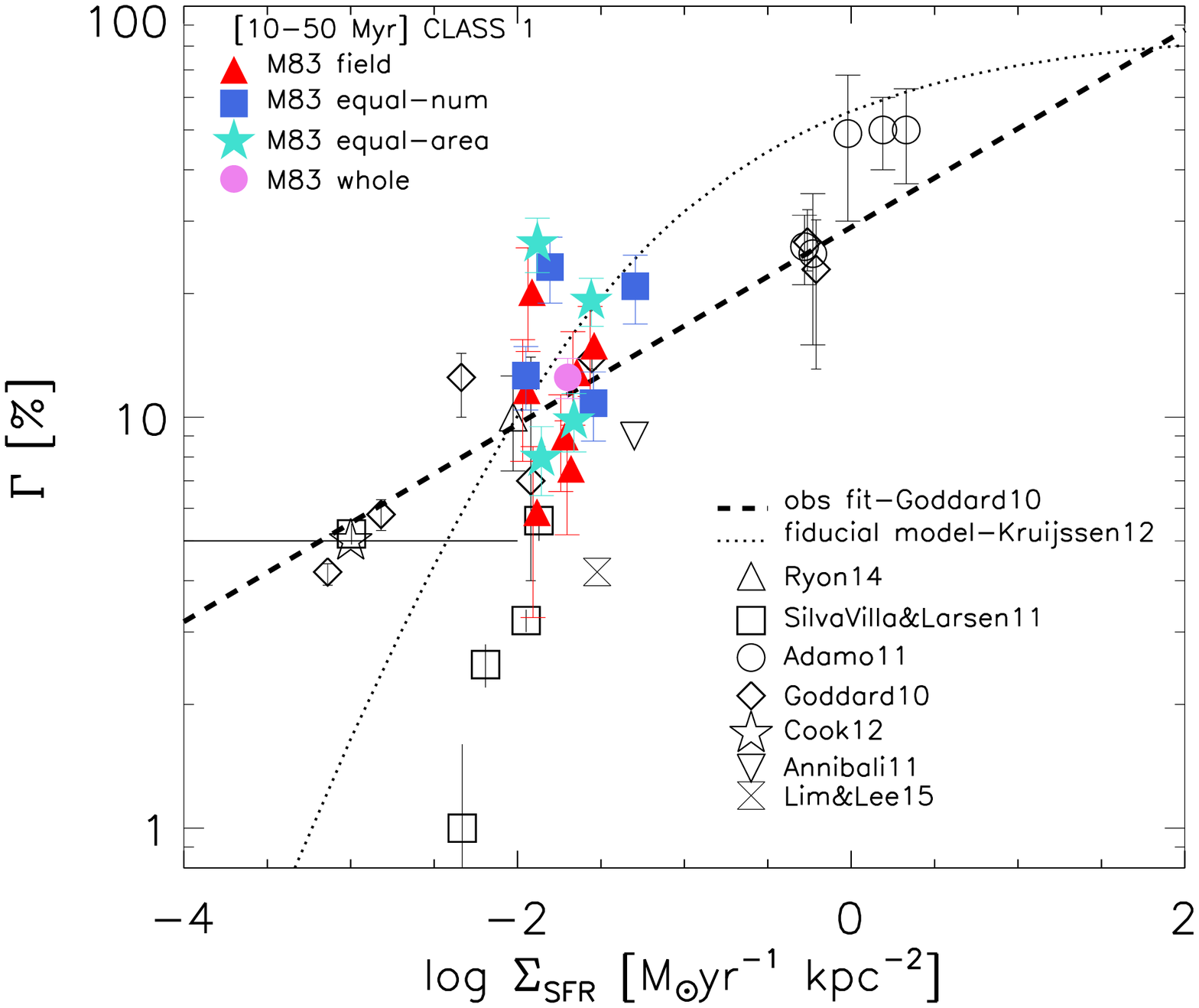}
\caption{\ga versus SFR surface density for class 1 clusters. The same diagram shows the position of the cluster formation efficiency estimated within different regions of the galaxy (see top inset) using circular bins of same area, annuli containing the same number of clusters, the position of each HST pointing (7 fields). On the left plot, \ga\, has been derived from clusters younger than 10 Myr, on the right plot, from clusters in the age range 10 to 50 Myr. Literature data and predictions have been included for comparison (see Table~\ref{tab5} in the Appendix for references and a complete list of the values used in this plot).}
\label{fig:gamma_sfr}
\end{figure*} 

\subsection{Cluster mass function as function of distance within the galaxy}
\label{icmf}
B12 reports evidence of a steepening of the ICMF in the cluster population of F2 relative to that of F1 (see the position of the two fields in the galaxy in Figure~\ref{fig:M83}). The two cluster populations are better fitted by a Schechter mass function of slope $-2$ and \mc$\sim 1.5 \times 10^5$ and $0.5 \times 10^5$ \msun\, in the inner and outer fields, respectively. Access to a complete cluster catalogue for this galaxy enables us to further  investigate how the galactic environment possibly influences the shape of the ICMF. Using the same analysis technique as in the previous section, we construct cumulative distributions of the cluster mass of each radial bin. Cumulative distributions are favoured to avoid any subjectivity in the choice of the mass bin size and because they are more sensitive to variations. Statistically, this method has the same significance of distributions of bins containing same number of objects \citep{2005ApJ...629..873M}, with our distributions containing only one object per bin.  In the previous paragraph, to estimate \ga\, we have limited our analysis to cluster ages overlapping the age range to which our chosen SFR traces are sensitive. However, to build the ICMF we do not need to apply the same limits. Similarly to the limits used by B12 and in other recent works, we apply a lower mass limit of 5000 \msun\, and an age cut between 3 and 100 Myr. For comparison, we generate Monte Carlo simulations of cluster populations with the same cluster numbers of the observations. We assume the same lower mass cutoff as in the observations (5000 \msun) and an upper limit to the cluster mass of $10^7$ \msun, that is much larger than the observed maximum cluster mass so it approximates the condition of no upper mass truncation of the ICMF.

We discuss below the results obtained using only class 1 clusters. In Figure~\ref{fig:mf_rad_app} of the appendix we include the same analysis but building the ICMFs of both class 1 and 2 objects, instead. The recovered absolute values vary within a factor of 2. Therefore, the observed trends and conclusions remain unchanged confirming that our results are not driven by the selection criteria.

\begin{figure*}
\centering
\includegraphics[width=5cm]{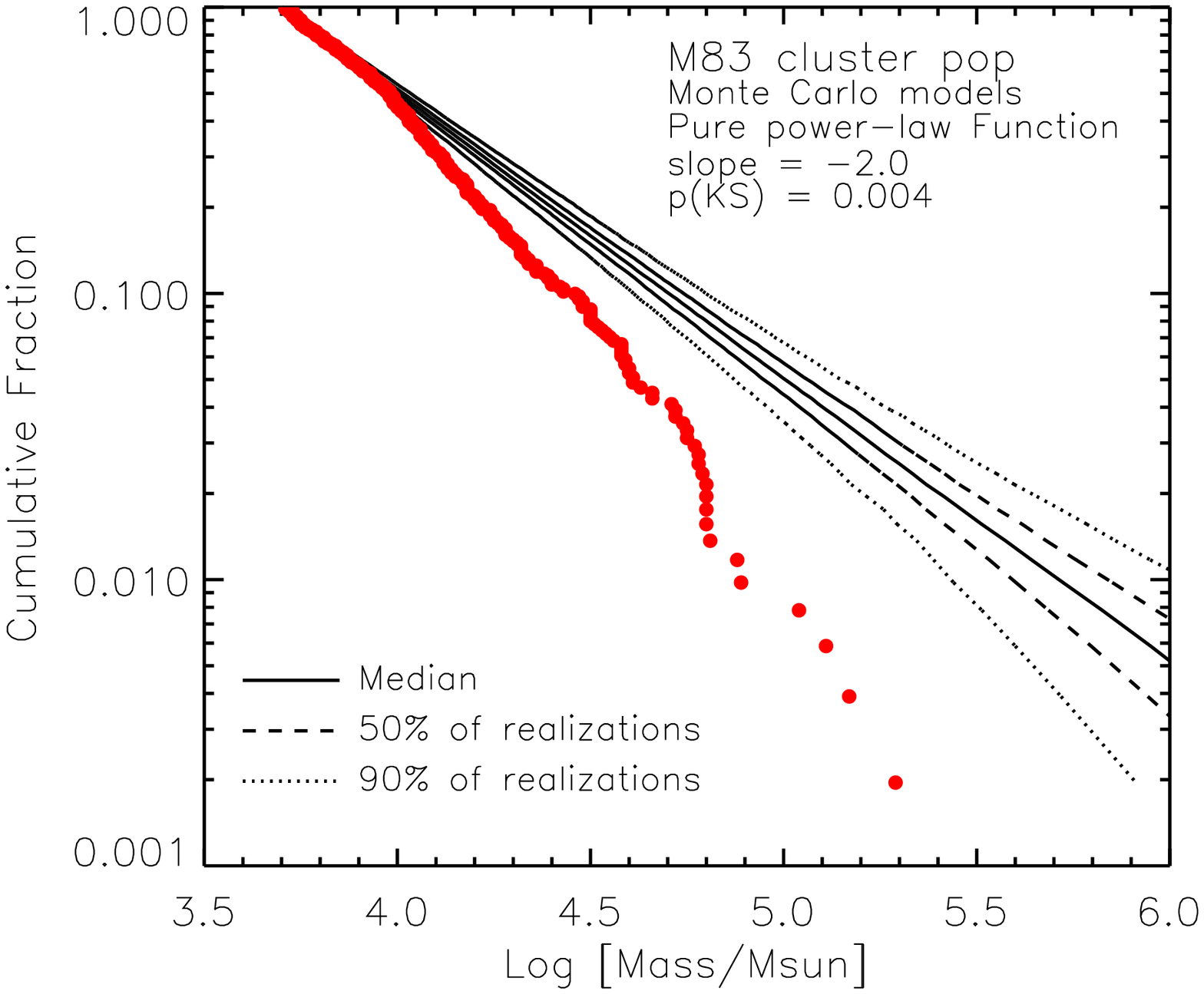}
\includegraphics[width=5cm]{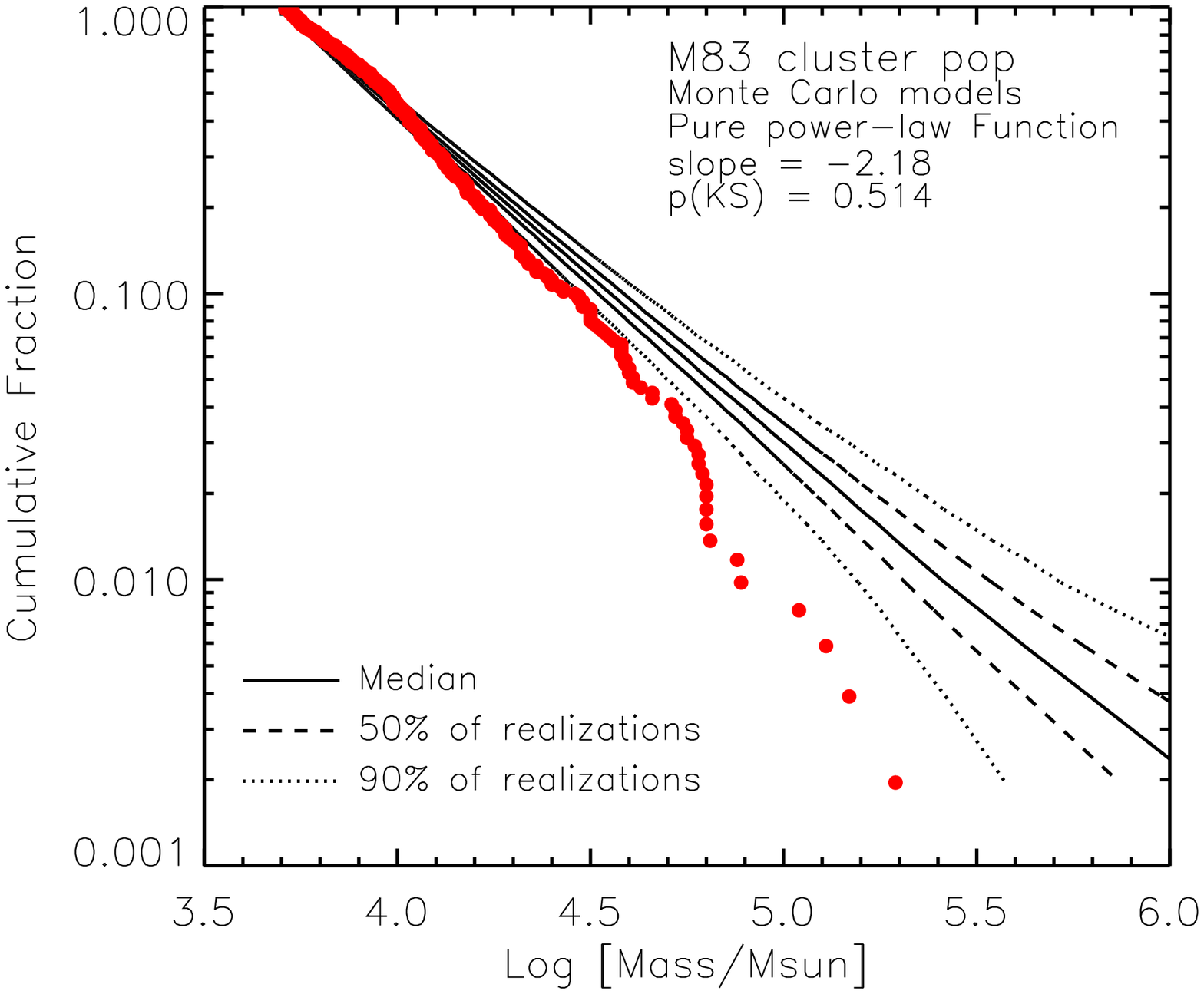}
\includegraphics[width=5cm]{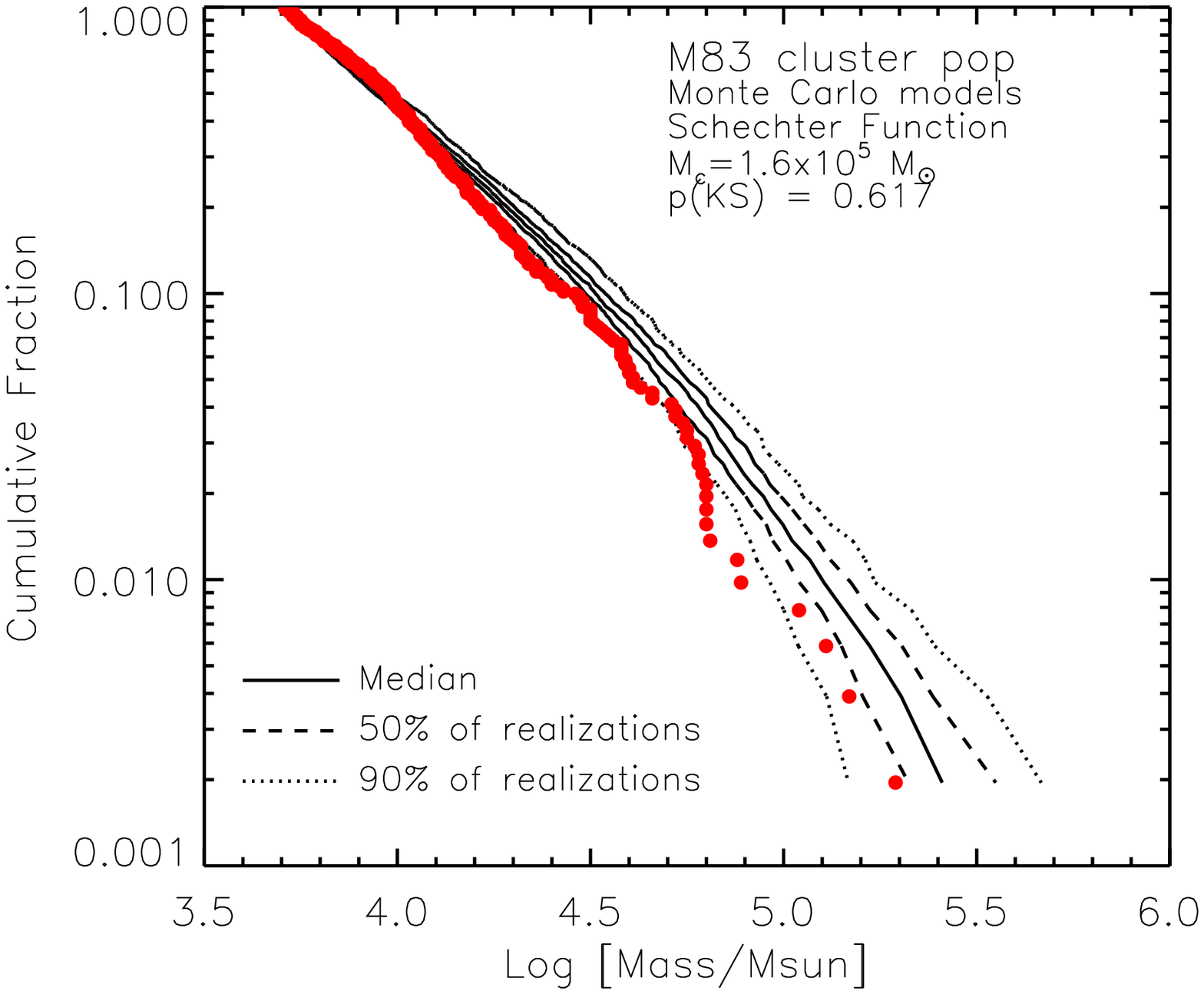}\\
\includegraphics[width=5cm]{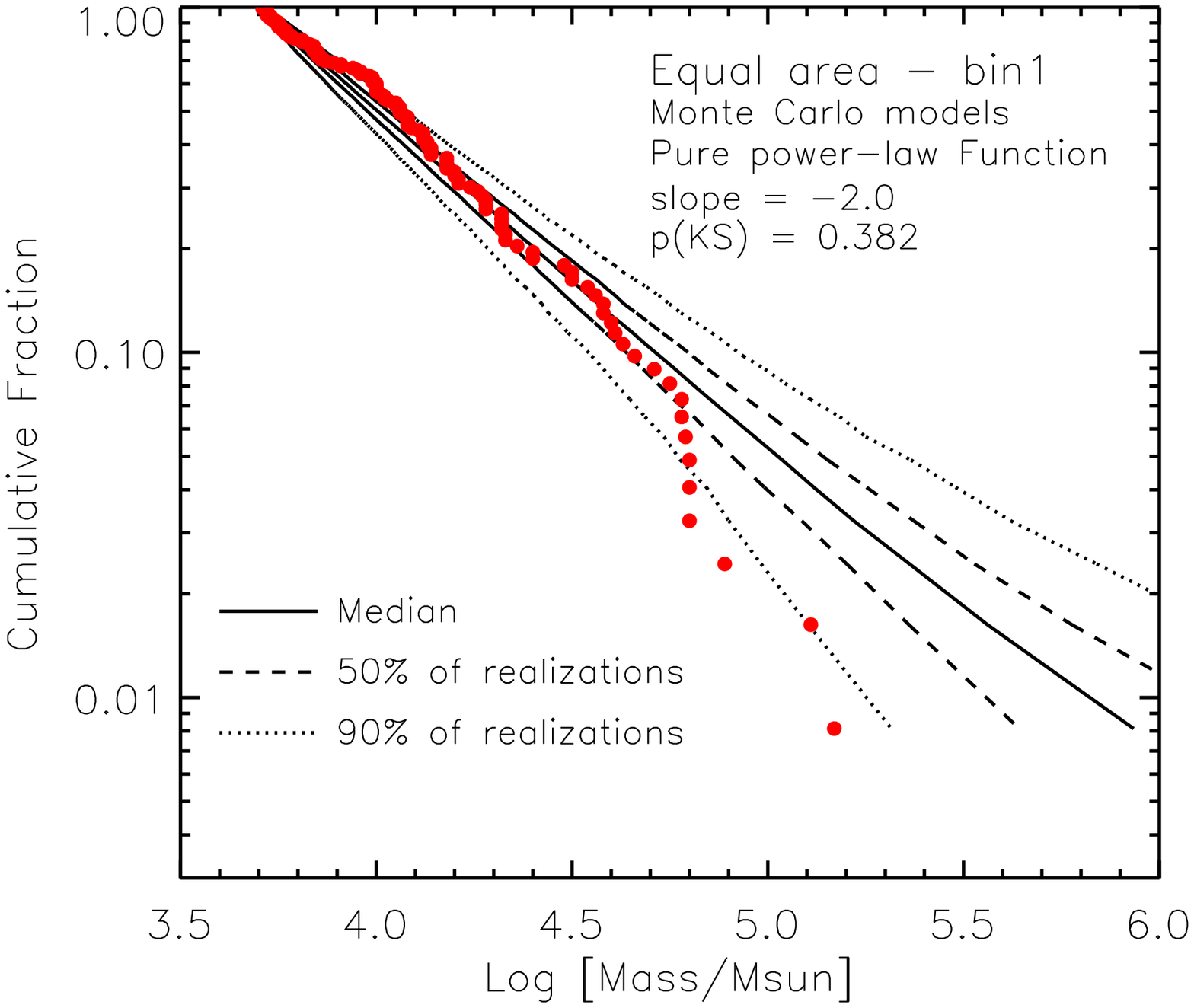}
\includegraphics[width=5cm]{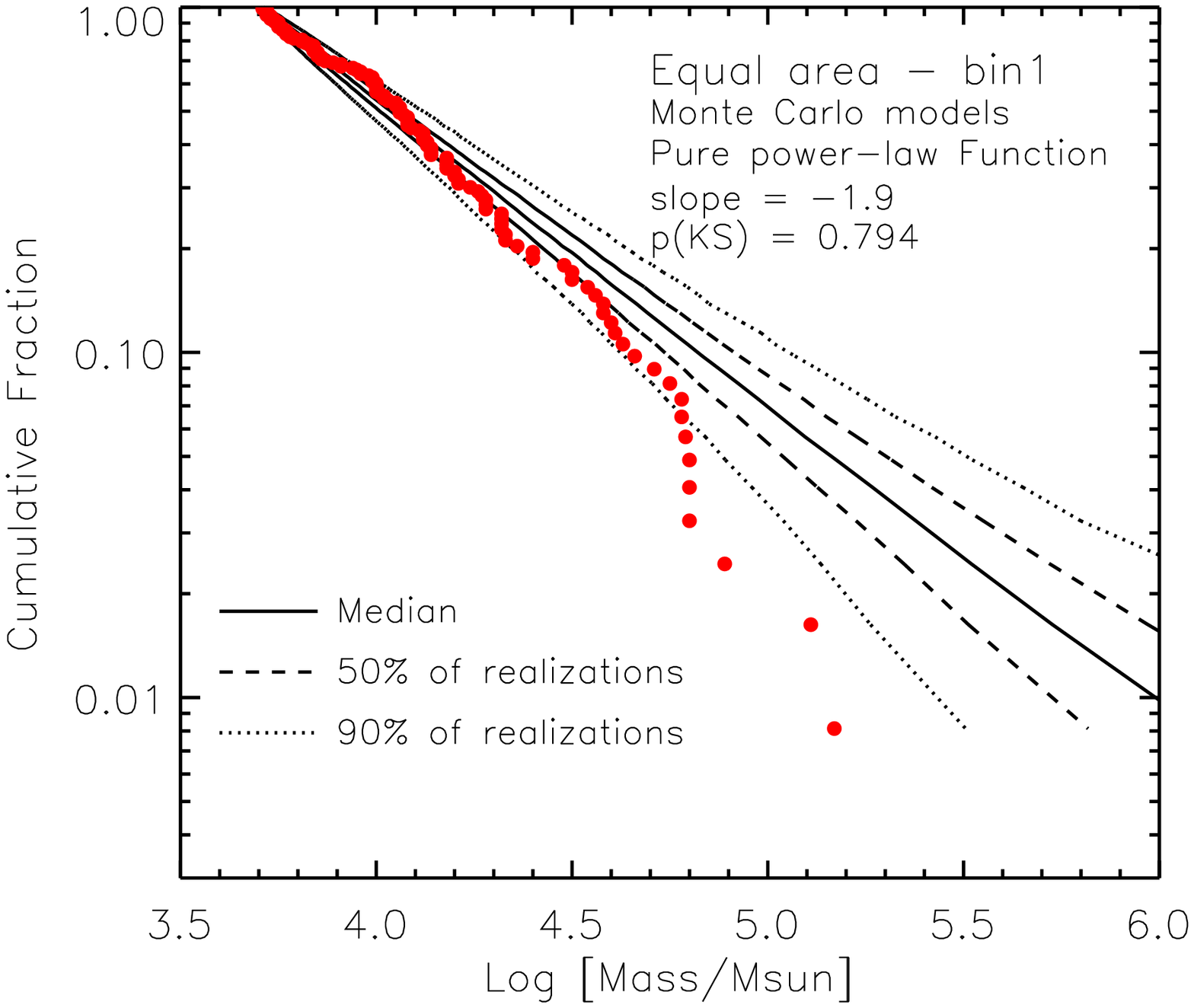}
\includegraphics[width=5cm]{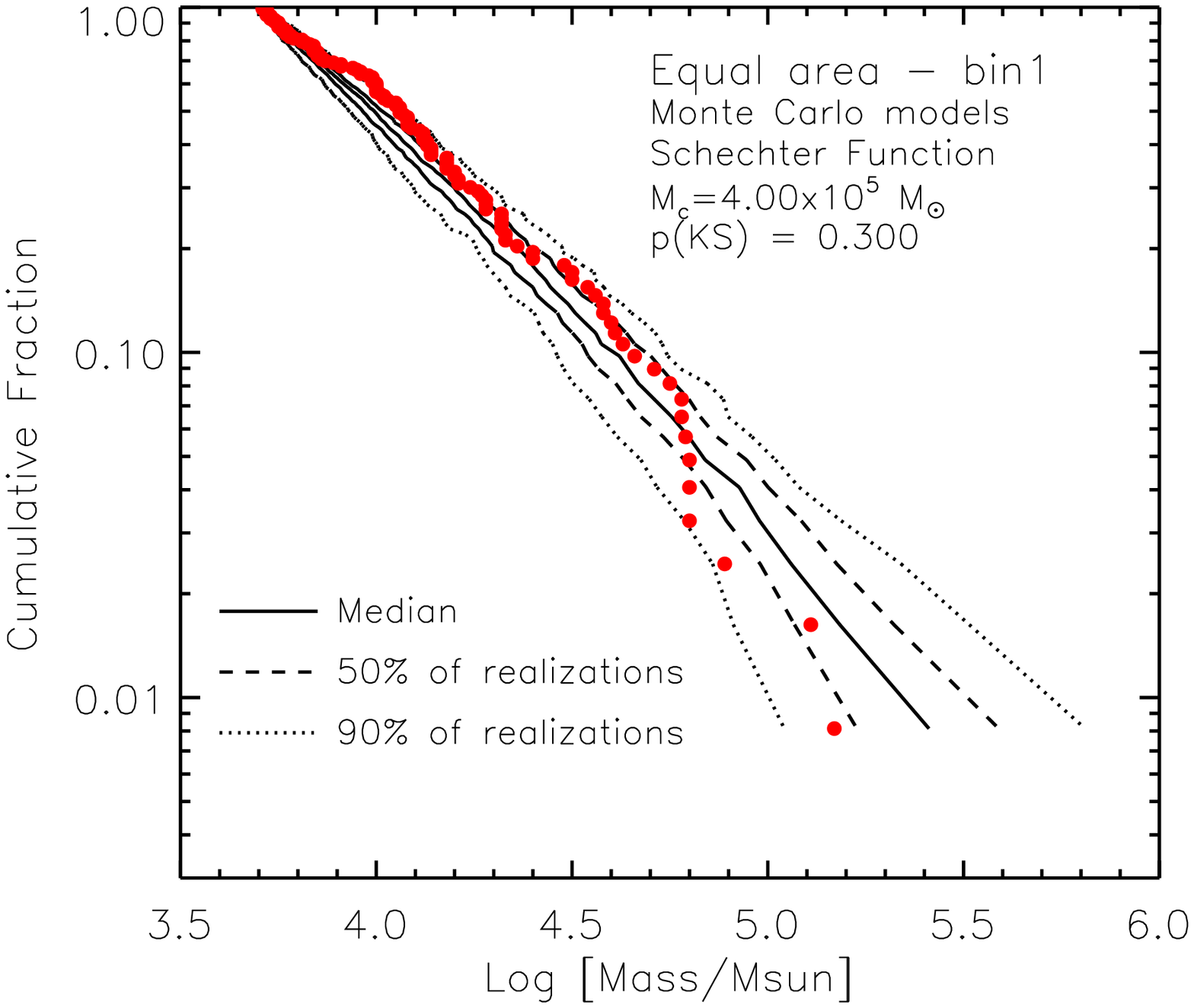}\\
\includegraphics[width=5cm]{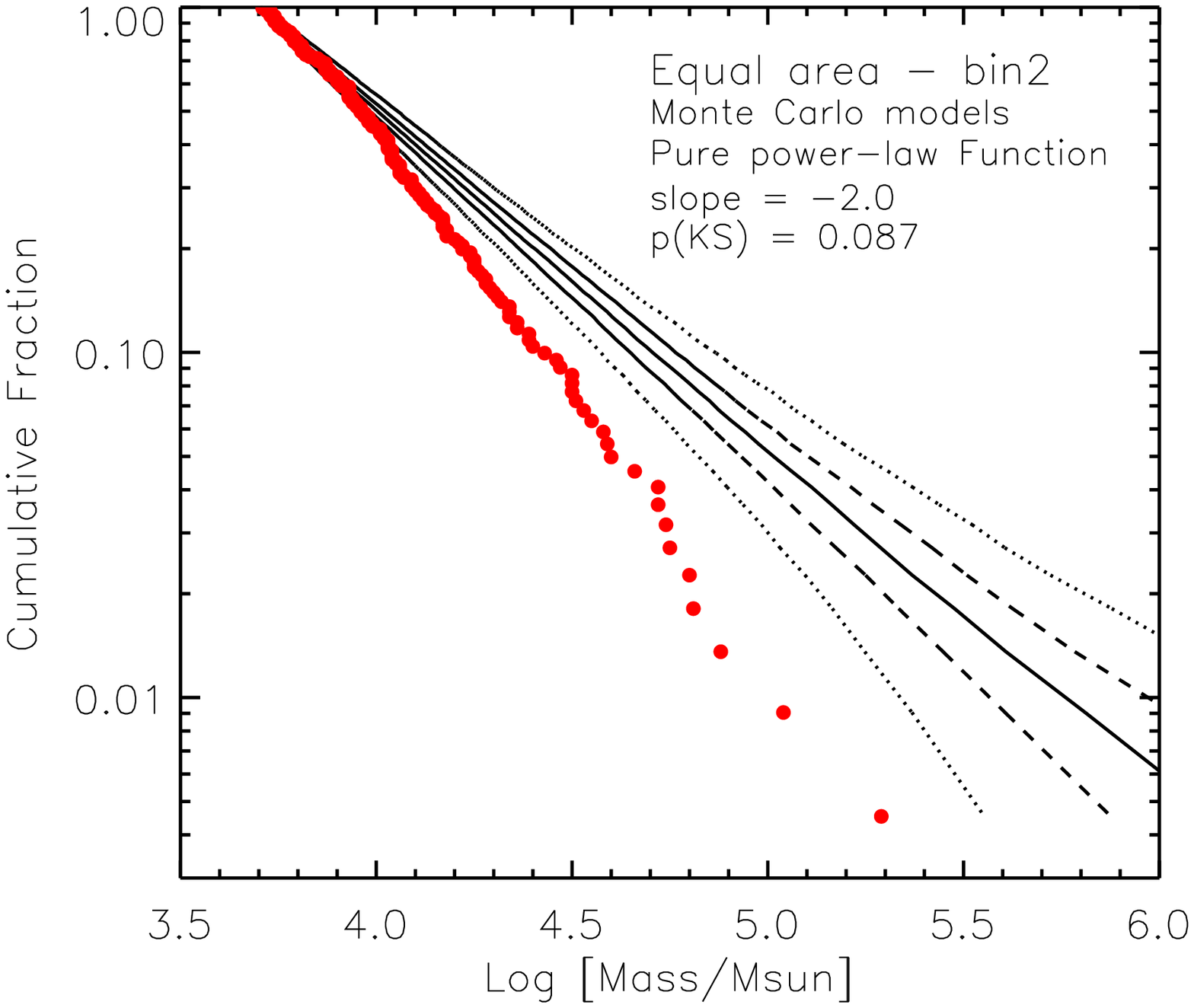}
\includegraphics[width=5cm]{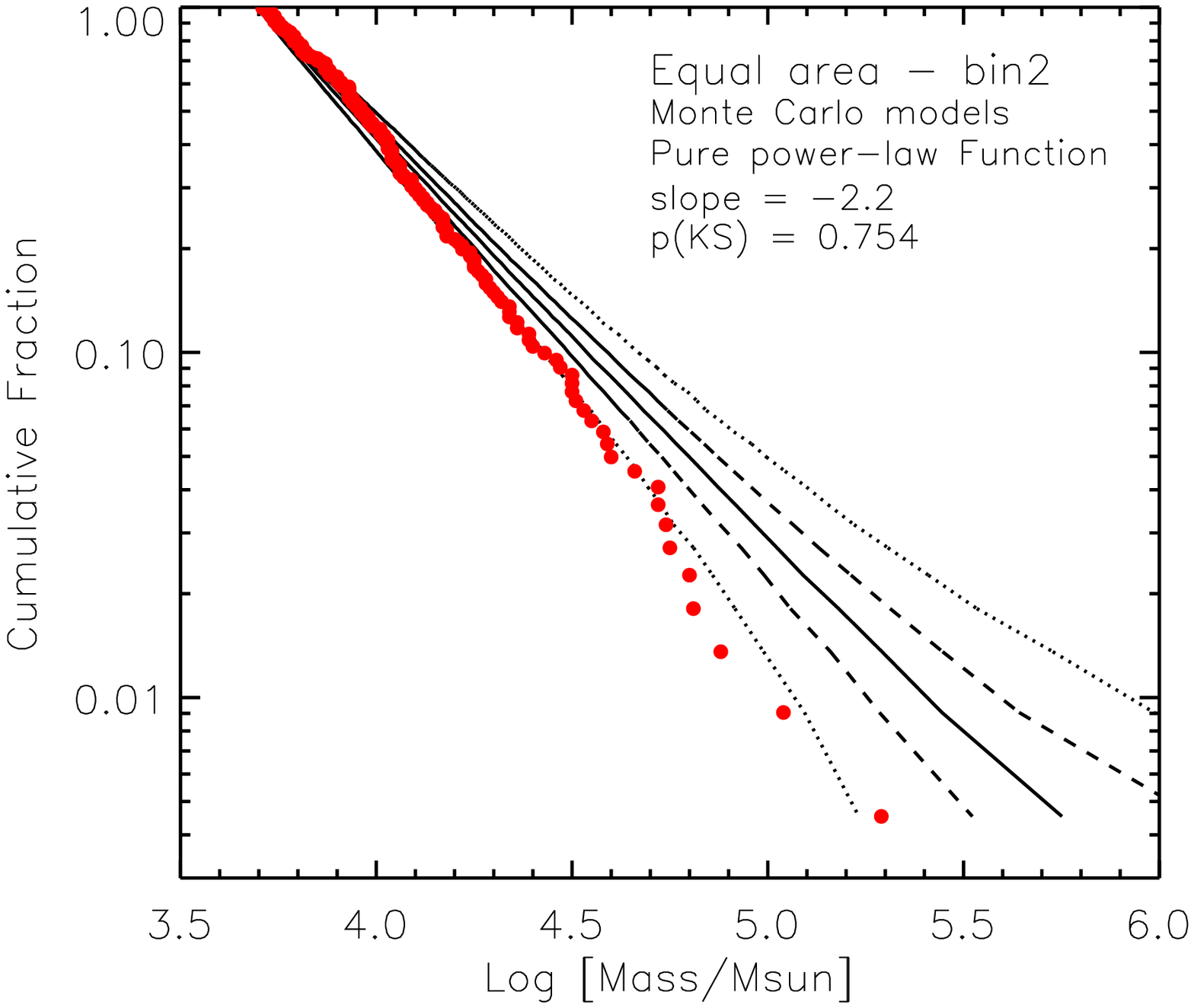}
\includegraphics[width=5cm]{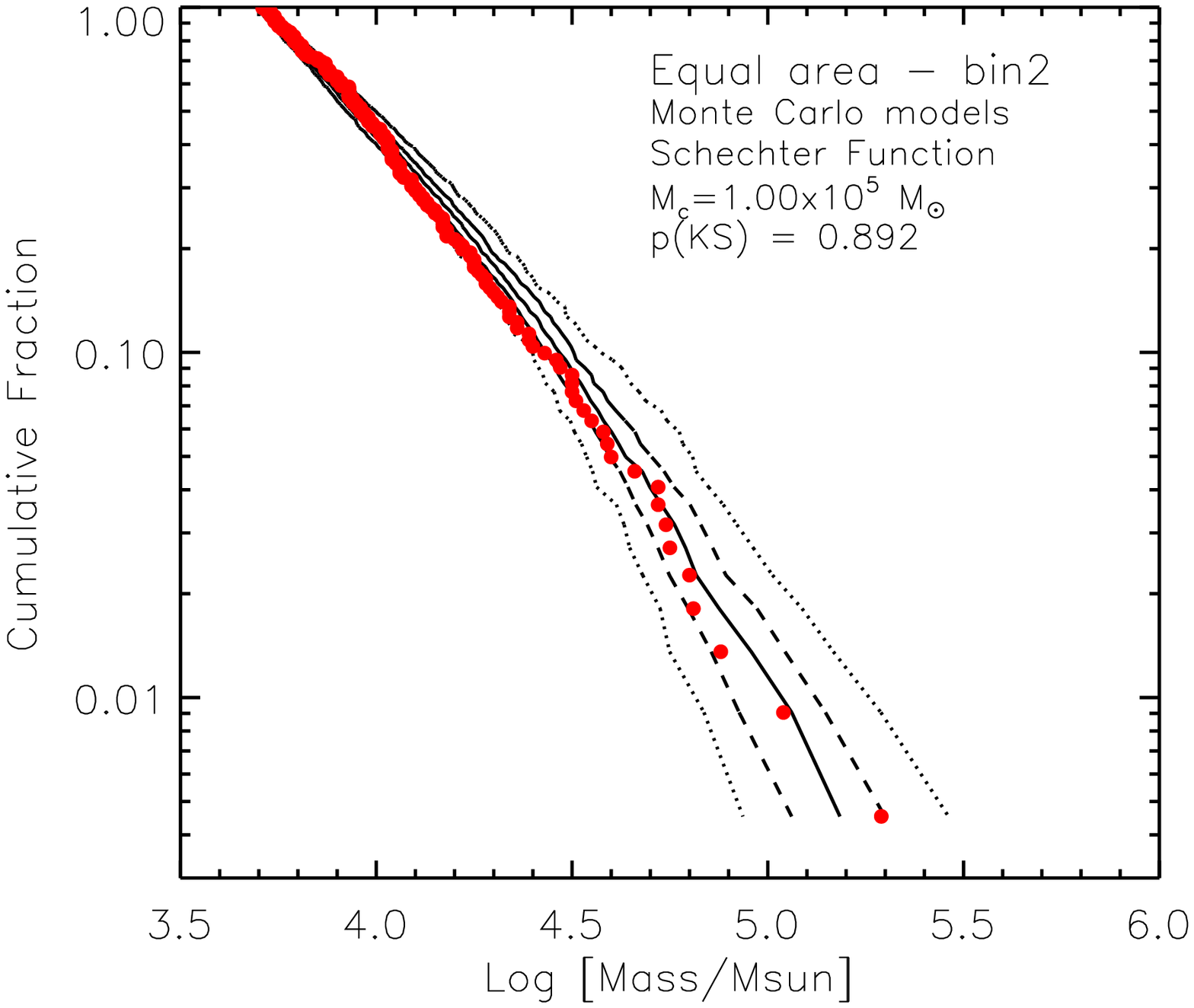}\\
\includegraphics[width=5cm]{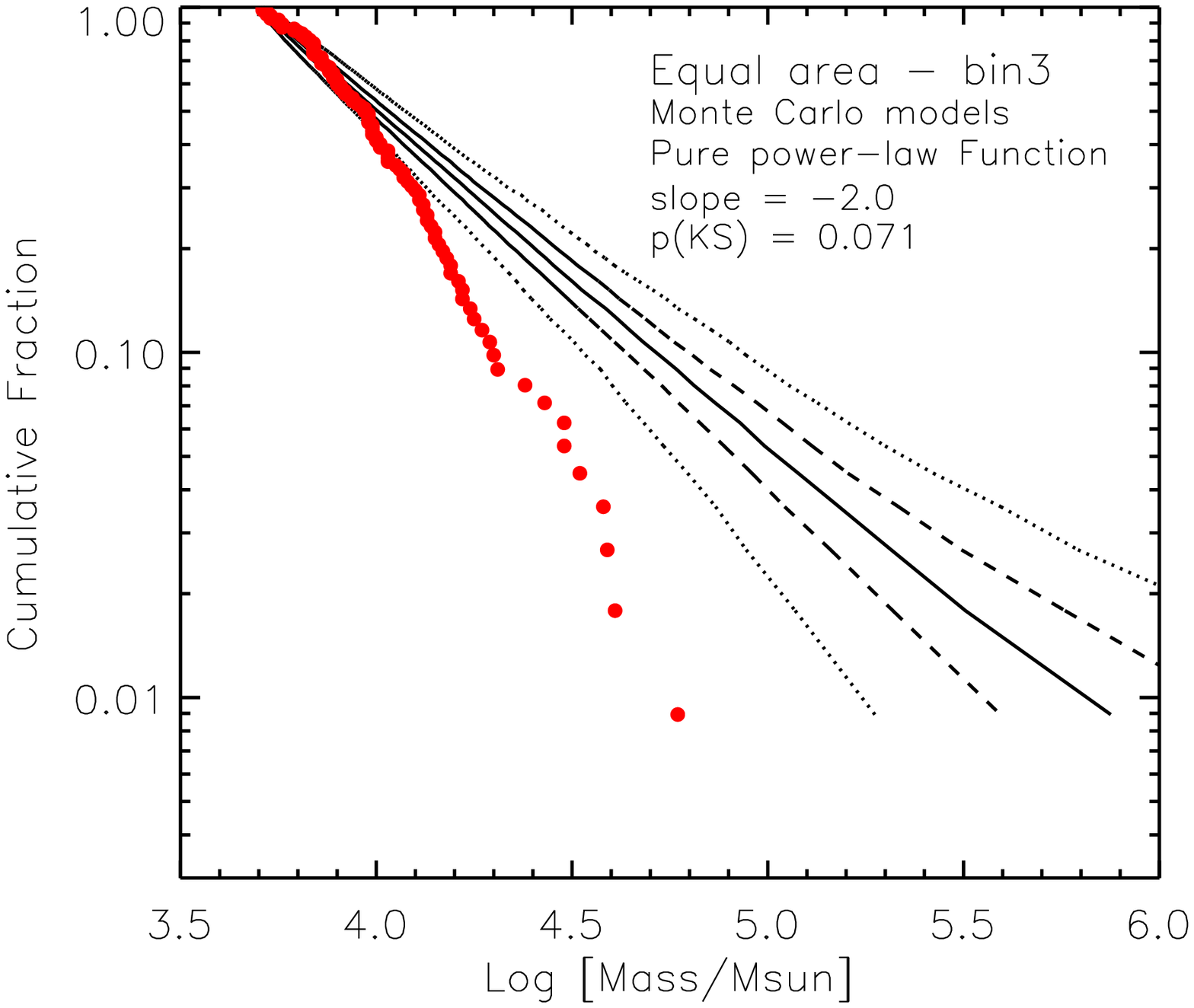}
\includegraphics[width=5cm]{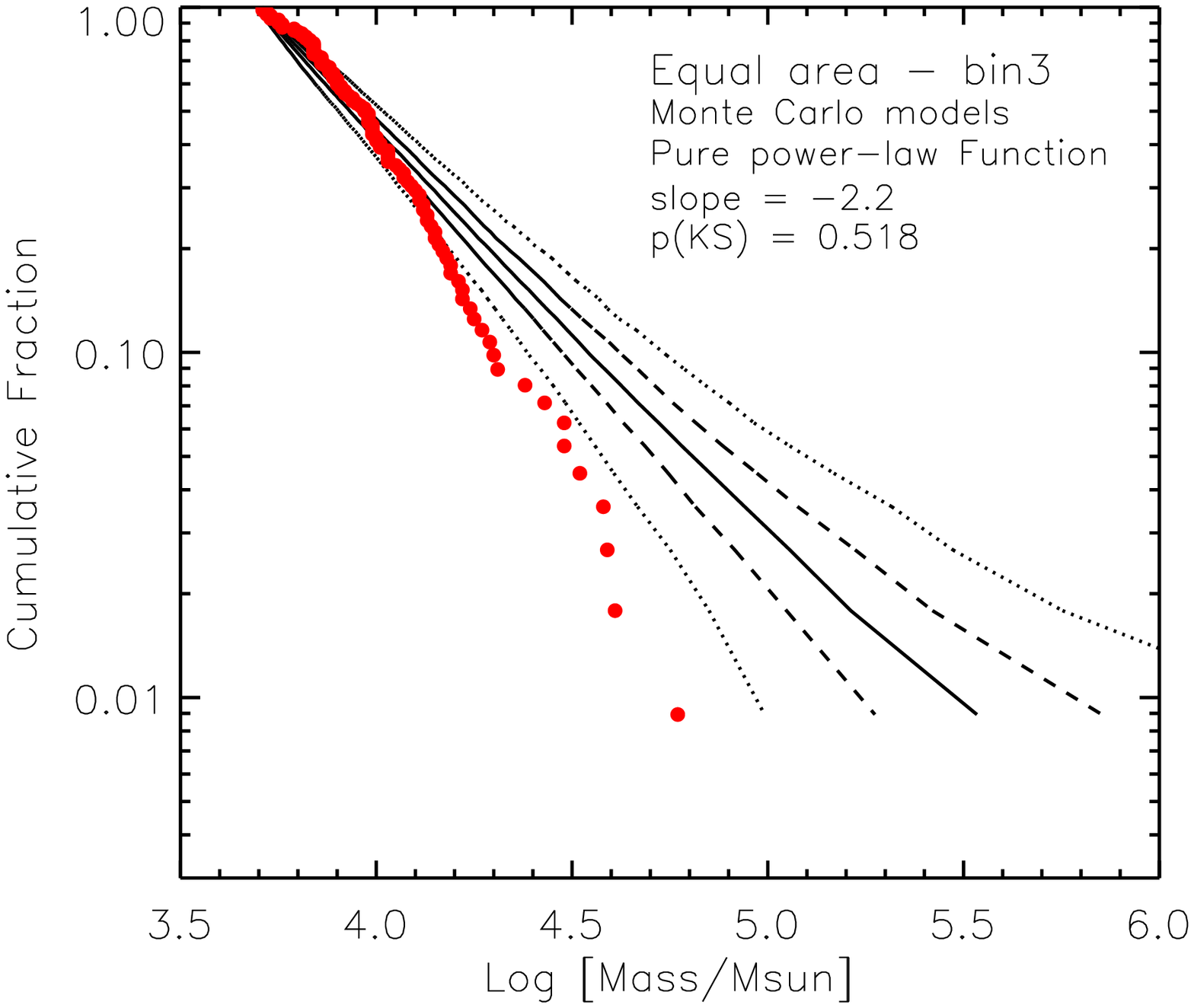}
\includegraphics[width=5cm]{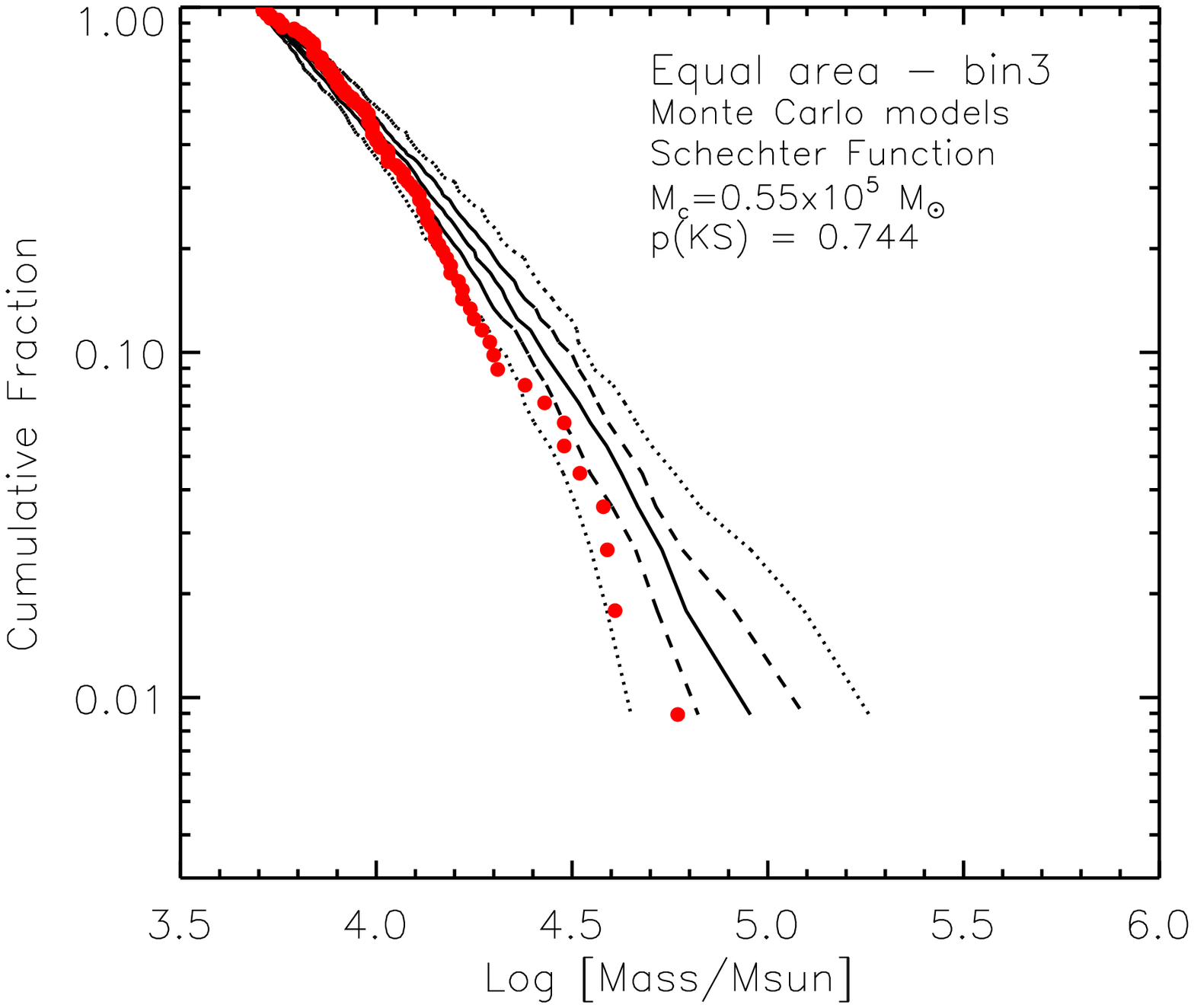}\\
\includegraphics[width=5cm]{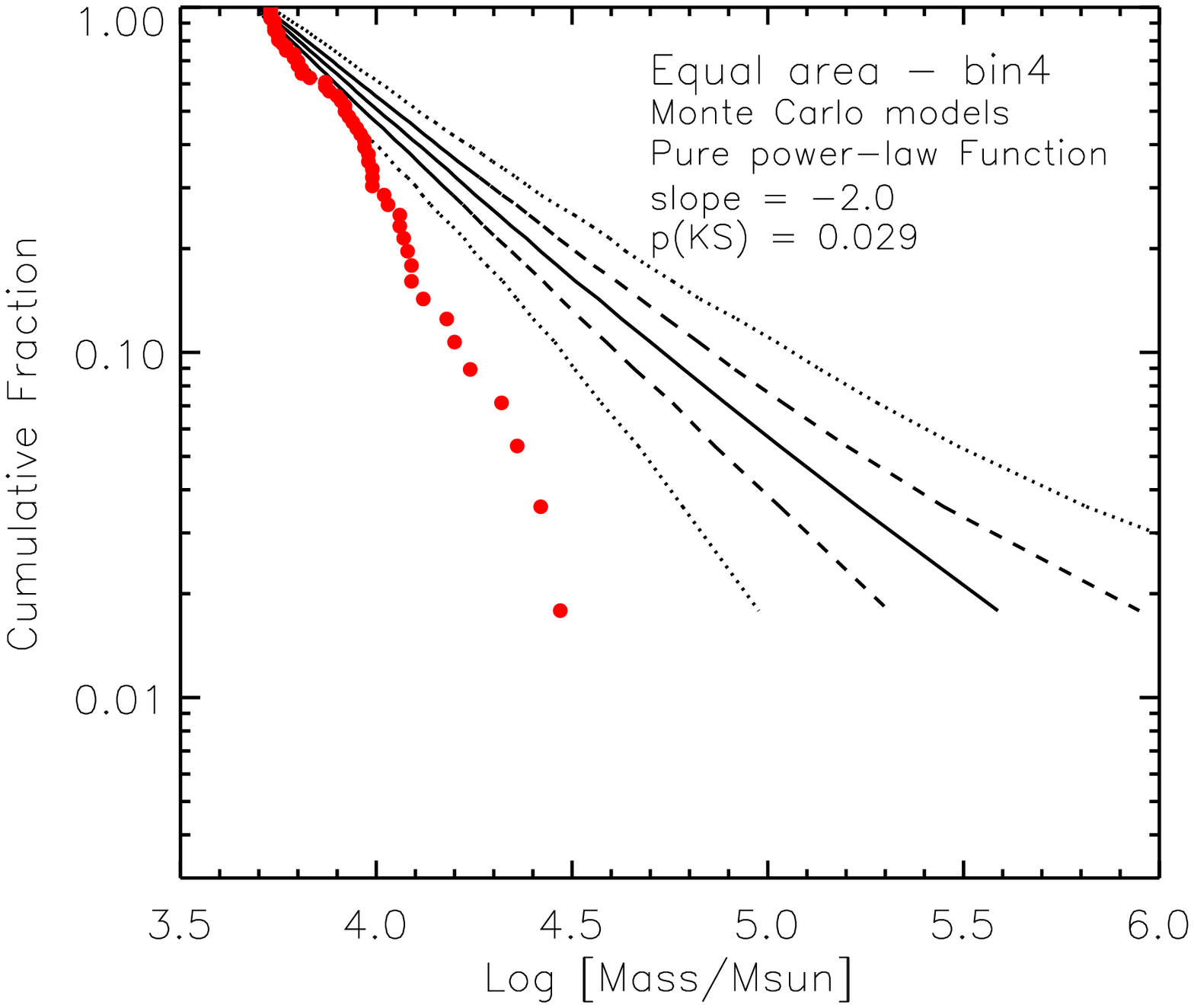}
\includegraphics[width=5cm]{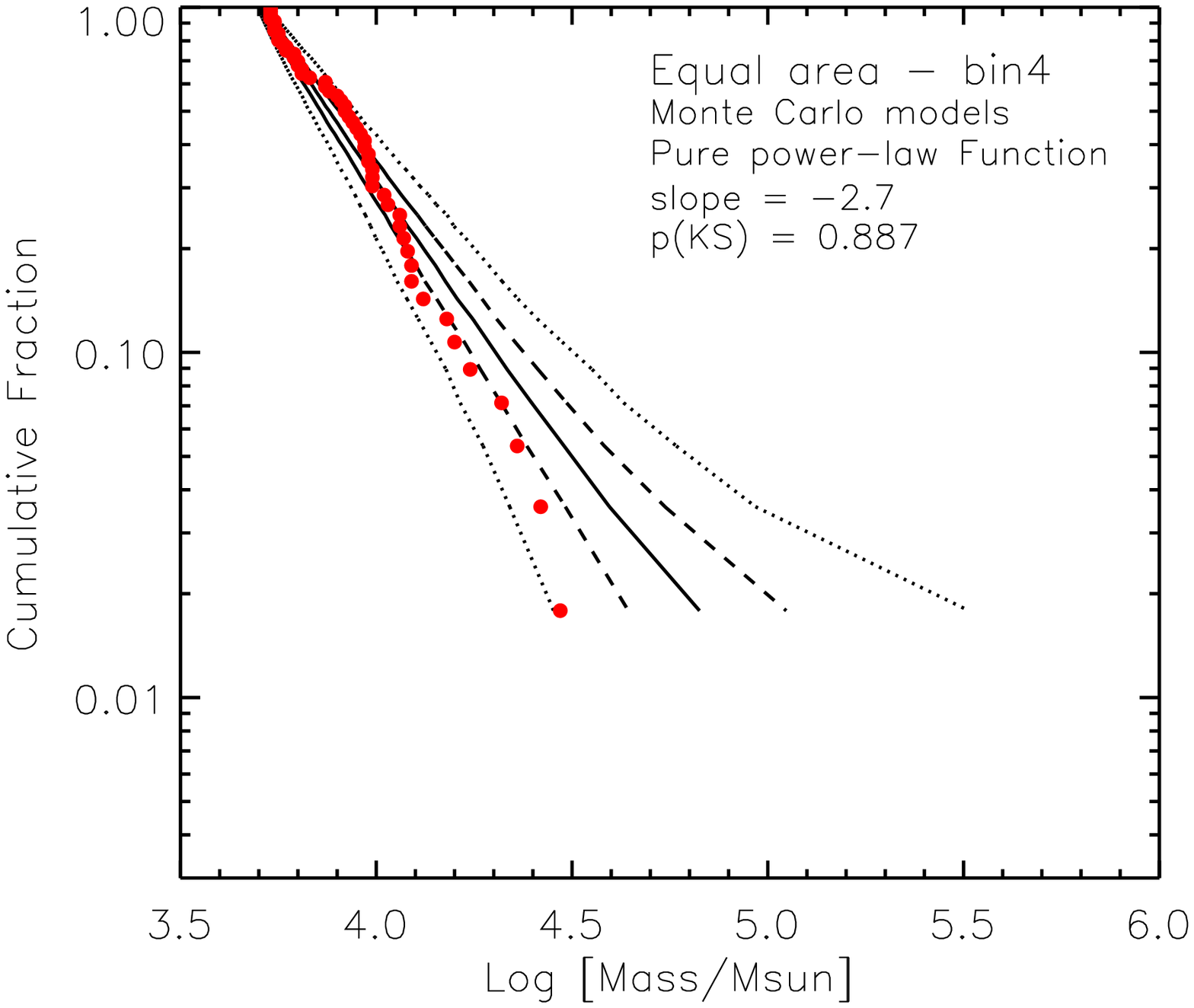}
\includegraphics[width=5cm]{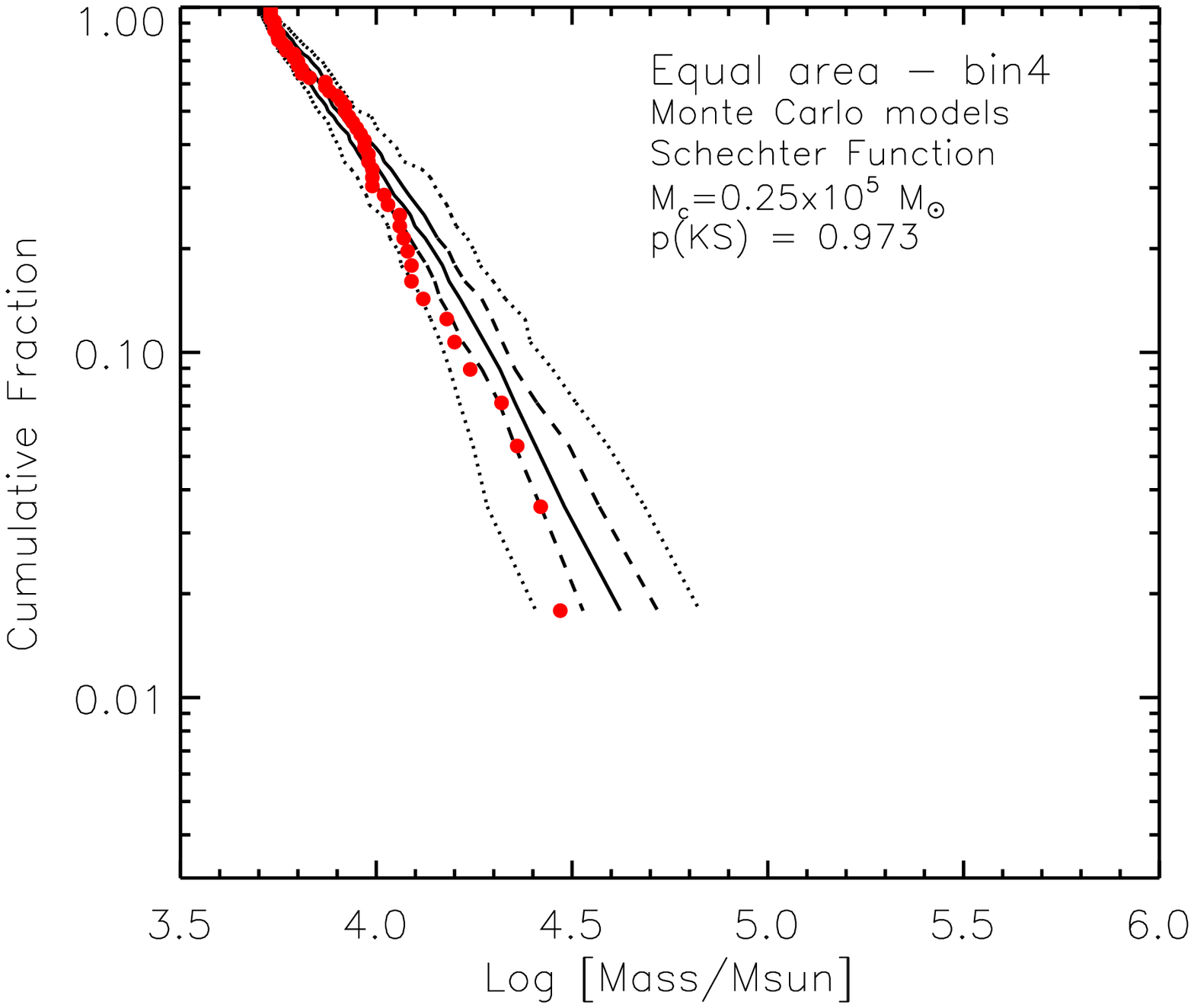}\\
\caption{Cumulative ICMFs (red filled dots) of the whole population and of the clusters within bins of same area (from top to bottom). Only clusters classified as 1 (see Figure~1 in the Appendix for the same plots but class 1 \& 2 instead), more massive than 5000 \msun\, and ages between 3 and 100 Myr have been included. Median (solid line), quartiles (dashed line) and extended boundaries (dotted line) of Monte Carlo simulations of cluster populations with the same number of objects as the observed ones are included in each panel. In the first column, we assume a pure power-law ICMF with slope $-2$. In the middle column we still assume a power-law ICMF but with varying slopes (see insets), chosen to maximise the Kolmogov-Smirnov probability that the observed and simulated distributions are drawn from the same parent distribution, p(KS). In the right column we assume that the ICMF is better described by a Schechter function of slope $-2$ and a exponential cut-off, \mc, chosen to maximise p(KS). }
\label{fig:mf_rad}
\end{figure*}

In Figure~\ref{fig:mf_rad}, we show the observed  and simulated ICMFs for the entire cluster population (top row) and for the clusters contained in bins of equal area (second top row to bottom). The results do not change if cluster populations in bins of equal numbers are used instead (see Table~\ref{tab2}). Each  panel shows the observed cumulative distribution (red dots), the median (black solid line) and the limits containing 50 (dashed lines) and 90 \% (dotted lines) of the 2000 cumulative ICMF realisations. We use different functions, i.e., single power law of slope $-2$ and no upper mass limit; single power law of some best-fitting slope that is not necessarily $-2$; a Schechter function of power-law slope $-2$ and exponential truncation above a variable characteristic mass, \mc. 

To verify the goodness of the agreement between the observed and simulated distributions we use the Kolmogorov-Smirnov statistic. The resulting probability, p(KS), that the two distributions are actually produced by the same parent population are included in each panel. In the first column we attempt to reproduce these distributions with a single power-law function of slope $-2$. The innermost bin cluster mass distribution (top left panel) is consistent with a power-law function of slope $-2$. The significance of the deviations from this function increases in the outer bins and the whole M\,83 cluster population (panels in the first column).  The middle column panels show an attempt to reproduce the ICMFs with single power laws of steeper slopes. We derive the value of the slope which maximises p(KS). Although the upper mass distributions are not well reproduced by a single power-law function, the KS test favours steeper slopes than the initial value $-2$.

The change in the slope from $\sim -1.9$ to $-2.7$ from the inner to the outermost bin is quite striking.  In Table~\ref{tab2} we report the uncertainties associated with the slope in each radial bin as well as for the whole galaxy. We derive the uncertainties on the slopes determining the index interval which contains 68\% of the Monte Carlo simulations for each cluster sample. As widely discussed in previous works \citep[e.g.,][]{2006A&A...446L...9G}, a steepening in the upper mass (luminosity) end of a distribution is often compatible with the presence of a truncation. In the third column we use a \citet{1976ApJ...203..297S} function, i.e. ${\rm d}N/{\rm d}M\propto M^\alpha\exp{(-M/M_c)}$, with slope $\alpha=-2$ and variable truncation mass \mc, as indicated in the inset of each plot (see also listed values in Table~\ref{tab2}). When using a Schechter function rather than a single power law slope $-2$, the agreement between simulations and observations improves for bin 2, bin 3, bin 4 and for the cluster population altogether. This shows that in 80\% of the cases, a Schechter function gives a substantially better match to the observed ICMF than a single power-law fit. A unique feature of the Schechter function is the ability to reproduce the observed distributions at the high-mass ends. Statistically a steeper power-law function and a Schechter type function give similar solutions, although the latter seems to be preferred everywhere but the innermost bin. Only there, the KS statistic favours a single power-law fit with a slope slightly less steep than $-2$, even though this slope cannot be ruled out within the uncertainties. For this reason, the Schechter fit produced with a power-law slope of index $-2$ does not improve the KS probability. If instead of using a Schechter function with slope $-2$ we use $-1.9$ the KS probability becomes p(KS)$=0.58$, which is still lower than the p(KS)=0.794 obtained for the single power-law fit. In section 3.4, we discuss the physics driving this difference.

Our results suggest that the ICMF changes significantly as function of distance from the centre and that the ICMF can be overall described by a power-law function of slope $-2$ and a decreasing \mc\, as function of distance from the centre of the galaxy. This finding is in agreement with the analysis presented by B12 for a smaller cluster sample in M83. It clearly suggests that the change in the star formation properties of the galaxy as function of radius is affecting the shape of the upper mass-end of the ICMF.

\subsection{The most massive cluster as function of distance within the galaxy}

\begin{figure}
\centering
\includegraphics[width=9cm]{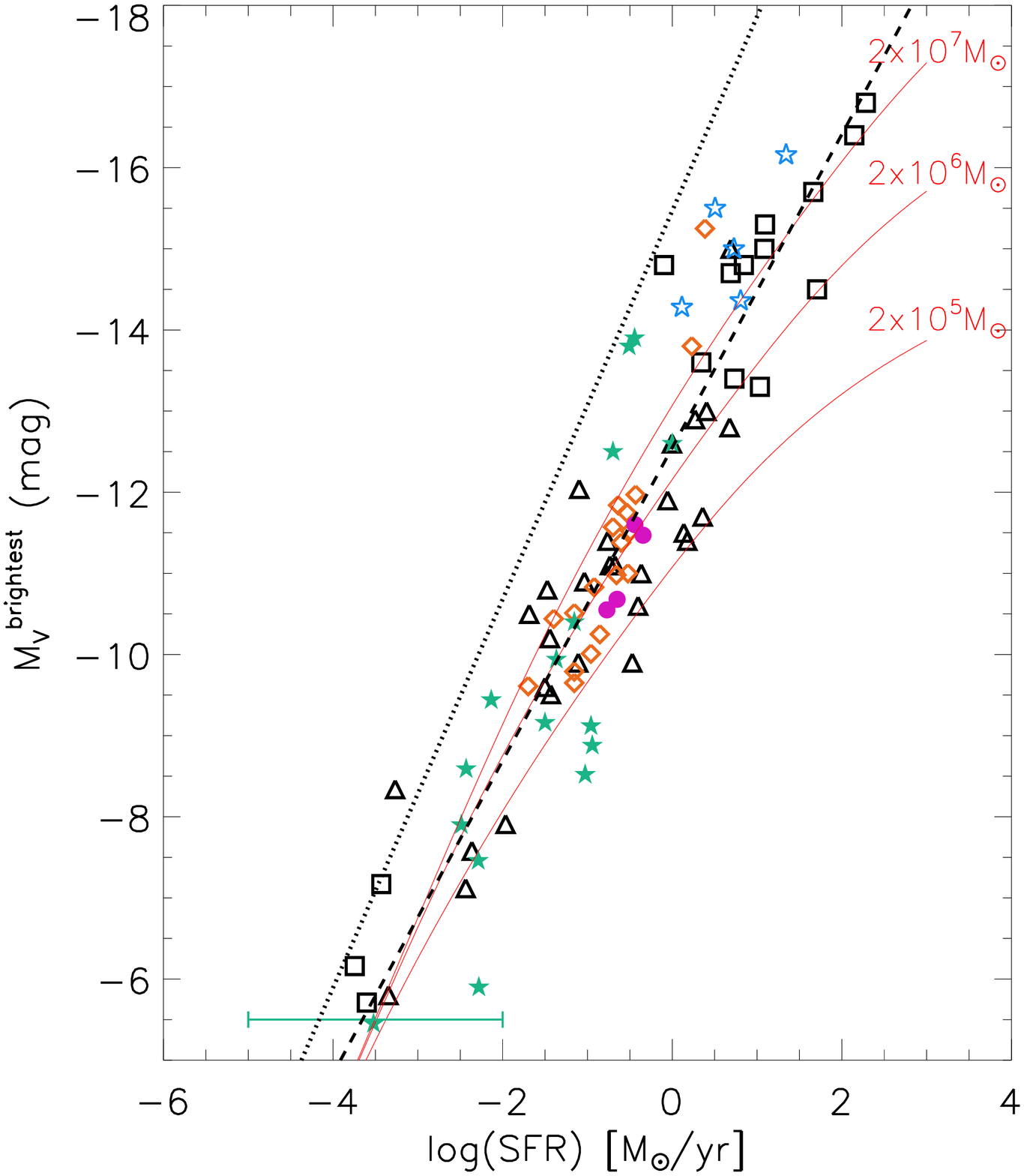}
\caption{The luminosity of the brightest cluster plotted against the star-formation rate (SFR) of the host galaxy. This plot contains a compilation of all data available in the literature. See Table~\ref{tab5} in the Appendix for a detailed description of the datasets. The dashed line is the best fit to the sample of galaxies plotted as triangles \citep{2002AJ....124.1393L}. The dwarf galaxy sample is represented by green filled stars \citep[see][for a description of the sample]{aa..nb..2015Spr}. Squares are the sample added by \citet{2008MNRAS.390..759B}, who derived the expected \relation relation if 100\% of stars are born in clusters with a power-law ICMF with index $-2$ (black dotted line). Blue stars are the sample of luminous blue compact galaxies studied by \citet{2011MNRAS.417.1904A}. The green horizontal bar shows the range of SFR of dwarfs which do not have young clusters \citep{2012ApJ...751..100C}. The M\,83 brightest young cluster of each the four bins of equal area are plotted as purple circles. The orange diamonds show the sample from \citet{2014AJ....147...78W}. The red solid lines show Monte Carlo simulations of cluster populations formed with a Schechter CMF with different \mc\, and a cluster formation efficiency of 8\%. }
\label{fig:mv_sfr}
\end{figure} 

Widely used in the cluster literature, the brightest cluster luminosity versus the SFR of the galaxy is a relation dominated by the size-of-sample effect \citep{2002AJ....124.1393L, 2008MNRAS.390..759B}. This relation, as well as the one presented by \citet{2000astro.ph.12546W}, shows that cluster formation is a stochastic process, i.e., the mass or luminosity of the most massive cluster in a galaxy increases with the SFR, because a larger number of formed clusters in the population increases the probability of forming more massive or luminous clusters. In Figure~\ref{fig:mv_sfr} we show the SFR versus M$_{V}^{\rm bright}$ relation including all the data available in the literature. We also add the brightest young cluster recovered in each of the 4 bins of M\,83 (purple dots). The dashed black line is a fit to the sample by \cite{2002AJ....124.1393L} and presented in \citet{2004MNRAS.350.1503W}. Using Monte Carlo simulations under the assumption that only a relative small fraction of star formation is happening in clusters (\ga$\sim8$ \%), \citet{2008MNRAS.390..759B} reproduces the observed trend. If all star formation would happen in clusters then the observed points should follow the dotted line on the left side of the dashed one. The scatter of the data points is quite large. Many factors can affect the location of cluster samples in this diagram, such as a recent change in the SFH of the galaxy \citep{2008MNRAS.390..759B}, or a difference in the \ga\, of each galaxy \citep{2011MNRAS.417.1904A}. 

Even though the SFR versus M$_V^{\rm bright}$ relation is mainly driven by the size-of-sample effect, i.e. the stochastic sampling of the most massive and luminous clusters from the ICMF, this does not rule out that the galactic environment still plays an important role in shaping the properties of the cluster population (see Section 4).  In Figure~\ref{fig:mv_sfr}, we include new Monte Carlo simulations of cluster populations formed with a Schechter ICMF of different \mc\, assuming a fixed cluster formation efficiency of 8\% (red solid lines). The presence of a truncation mass in the ICMF clearly introduces a bend in the simulated SFR versus M$_V^{\rm bright}$ relation. However, this bend is absent in the observed SFR versus M$_V^{\rm bright}$ relation, because galaxies with higher SFRs tend to have higher truncation masses $M_c$ (see below). 

\begin{figure}
\centering
\includegraphics[width=9cm]{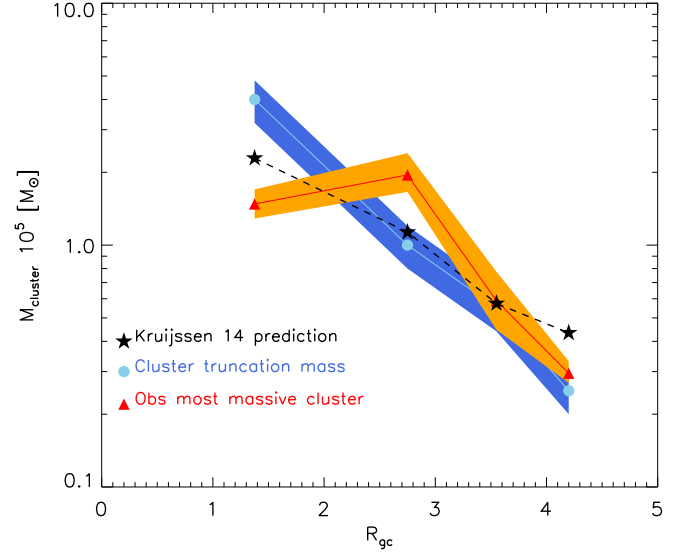}
\caption{The mass of the most massive cluster found in each bin of same area is plotted as function of galactocentric distance (red triangles). The shadowed area in orange show the uncertainties on the mass estimates. The \mc\, derived in Figure~\ref{fig:mf_rad} are plotted using cyan dots, while the blue band show 20\% tolerance margin for the derived values. The predicted cluster \mt\, are included as black stars (see the text and \citealt{2014CQGra..31x4006K}).}
\label{fig:mmax_rad}
\end{figure} 

The presence of a varying upper mass cutoff in the ICMF is supported by observations of GMCs in local galaxies. In the M\,51 galaxy, \citet{2014ApJ...784....3C} report clear differences in the number, mass, luminosity, density, and mass distributions of GMCs within different environments of the galaxy (e.g. inter-arm and arm regions, molecular ring, etc.). The mass of the most massive GMC in M\,51 changes as function of the galactic environment. With a similar argument, \citet{2009A&A...494..539L} noticed that the maximum GMC mass observed in the Antennae system is larger than the GMC mass found in local spiral galaxies. Clusters form within GMCs, and their final mass is a fraction of the total mass of the GMC. The total fraction of gas mass converted into stars is usually described by the star formation efficiency, which observations of galactic star-forming regions suggest is only of a few per cents \citep[e.g.,][]{2002ApJ...577..206E, 2009ApJS..181..321E}. The maximum fraction of the GMC mass that can end up in a single, most massive cluster then requires this star formation efficiency to be multiplied by the CFE \citep{2014CQGra..31x4006K}. The observation of an ICMF truncation mass at the high-mass end is possibly linked to the limit imposed by the galactic environment to the maximum mass that GMCs can have. Theory predicts that this mass-scale should correspond to the maximum mass able to overcome the galactic differential rotation and undergo collapse (i.e. the \citealt{toomre64} mass, see \citealt{2014CQGra..31x4006K} and references therein). 

In Figure~\ref{fig:mmax_rad}, we show the observed maximum cluster mass within each bin in M\,83 (red triangles and mass uncertainties derived from SED fitting), the recovered \mc\, including 20\% uncertainties, and predicted maximum cluster mass-scale using equation 3 from \citet{2014CQGra..31x4006K} (black symbols). It is important to stress here that although we refer to \mc\, as a truncation it is not a sharp limit. The truncation of a Schechter function reflects an exponential decay of the probability distribution at the high-mass end. The chances of having clusters more massive that \mc\, are lower than if a pure power-law function of slope $-2$ are used but not impossible \citep{2009A&A...494..539L}. Indeed, we see that the observed M$_{max}$ and the recovered \mc\, are in agreement within a factor of 2. To derive the predicted maximum cluster mass we use the averaged \ga\, (mean of the values at 1-10 and 10-50  Myr ) observed in each bin of equal area and we assume that the GMC mass changes  between $2$ and $0.8\times10^7$ \msun\footnote{We prefer using the observed maximum cloud masses rather than calculating the Toomre mass using the velocity dispersion profiles from \citet{2004A&A...422..865L}. These authors discuss that beam smearing yields velocity dispersions which (especially at small galactocentric radii of $<2$~kpc) can be overestimated by a factor of 2. This would subsequently lead to severely overestimated Toomre masses ($M_{\rm T}\propto \sigma^4$) by up to an order of magnitude.} \citep{2004A&A...422..865L} in each bin and fix the star formation efficiency to 5\%. In spite of the approximations, the predicted and the observed trend agree remarkably well, supporting the role of the galactic environment in shaping the upper mass limits of the ICMF.

\section{Discussion} 
\label{sec:discussion}


In this work, we probe the impact of the environment on the cluster formation process. Our analysis reinforces previous findings and presents unambiguous imprints left by the M\,83 galactic environment on its own cluster population.

Cluster formation is a stochastic process, however the efficiency of the process, i.e. the amount of star formation locked in bound star clusters, varies as function of the $\Sigma_{\rm SFR}$. Several works have shown this link on galactic scales \citep[e.g.,][among many others]{2010MNRAS.405..857G, 2011MNRAS.417.1904A}. This effect is also observed on sub galactic scales, when looking on different regions of the same galaxy \citep[e.g., SV13,][]{2014AJ....148...33R}, and our new analysis confirms this result.  We find that the cluster formation process is tightly linked to the gas surface density, just like the SFR density is via the Schmidt-Kennicutt relation \citep{1959ApJ...129..243S, 1998ApJ...498..541K}. As the gas pressure increases (and hence the gas or SFR surface density) the density spectrum of the ISM broadens and a larger fraction of the density fluctuations reaches densities high enough to collapse on short time-scales and achieve high star formation efficiencies. This locally more efficient star formation enables the young stellar structure to remain bound when any residual gas is expelled \citep{2012MNRAS.426.3008K}. As a result, the CFE increases with the gas pressure, gas surface density, and through the Schmidt-Kennicutt relation also with the SFR density.

However, \ga\, appears to not grow indefinitely as it obviously cannot attain values exceeding 100\%. At gas densities above $\Sigma(H_2)>1000$ \msun/pc$^{-2}$ (which corresponds to \ga$\sim$70 \%) \ga\, will not increase significantly \citep[see dotted line in Figure~\ref{fig:gamma_sfr};][]{2012MNRAS.426.3008K}. Denser gas favours star and cluster formation but at the same time destroys low density systems more easily via GMC encounters. These tidal perturbations imply that a CFE of 100\% is never reached.

The presence or absence of a truncation at the high-mass end of the ICMF has been long debated in the literature. With simple numerical simulations \citet{2006astro.ph..6625L} shows why it is so challenging to prove the presence of a truncations. For example, in Section~\ref{icmf}, we find that the ICMF of the innermost bin can be well described by a single power-law mass function of slope $-1.90\pm0.11$. However, the presence of a truncation cannot be ruled out from the KS test. Large cluster populations are necessary to well sample the high mass end of the ICMF, but in local galaxies cluster populations have usually a few hundred objects  \citep{2006astro.ph..6625L}. Some previous works in the past have found evidence of a possible truncation and/or variation of the ICMF within the same galaxy \citep[e.g.,][]{2009A&A...494..539L, 2012MNRAS.419.2606B, 2013AJ....145..137K} or steepening of the cluster luminosity function \citep[recently][]{2014AJ....147...78W}. In our work we observe a clear steepening of the ICMF, well represented by a Schechter type distribution of slope $-2$, and a variable \mc\, as function of distance from the centre of the galaxy. The Schechter function  combines the power-law slope and the presence of a limit above which the probability to form more massive clusters goes exponentially to zero. It is not a sharp limit. Interestingly, we observe a clear decline of \mc\ within the same galaxy. As for the CFE, the decrease of the truncation mass is most likely related to the gradient of the gas pressure (and hence gas and SFR surface density). The galactic environment thus poses a direct limit on the maximum possible mass-scales of GMCs and stellar clusters.

The globular cluster mass function has been reported to be well fitted by an evolved Schechter function by \citet{2007ApJS..171..101J}. This function takes into account the stellar mass loss suffered by globular clusters at the low-mass end of the mass function as well as the steep, non-gaussian decline at the high-mass end. The \mc\, of each globular cluster population increases with the total $B$ band luminosity (mass) of the host galaxy. Dynamical friction alone is not able to reproduce the observed decline of \mc\, for lower luminosity galaxies suggesting a link between \mc\, and the host galaxy at the moment the globular cluster mass function was created.  

The results reported in our analysis of M\,83 and supported by evidence found in previous works can shed light on the cluster formation process not only in local galaxies but also at high redshift. The young star cluster mass function can be described by an almost universal ICMF of slope $-2$ and a variable \mc\, which appears to be a function of the host galaxy environment, in that it increases with the gas pressure. Combining this with the aforementioned increase of the truncation mass of the globular cluster mass function with the host galaxy mass, we infer that the gas pressure at the time of globular cluster formation likely increased with the host galaxy mass, as is indeed proposed by recent models \citep{Diederik15}. Our results can help to understand the physical process under which globular clusters formed. Both young and ancient cluster populations can potentially be used as tracers of the star formation process of their host galaxies. However, it is important to keep in mind that the physical conditions of the ISM where globular clusters have formed are quite different and  nowadays observed only in merging starburst systems \citep[e.g.,][]{1992AJ....103..691H, 1993AJ....106.1354W}. Moreover globular cluster populations have most likely survived because during merging events they have been relocated in the halos of their galaxies, thereby escaping the gas-rich, disruptive bodies of their host galaxies \citep{Diederik15}. For instance, the M83 cluster population will be affected by cluster disruption according to the results of SV14. In the absence of major events which could move clusters out of the disk, this population will not survive for a Hubble time.

\section{Conclusions}

The aim of this work is to probe the link between cluster formation, the observed statistical cluster properties and the galactic environment where they form and interact. With respect to previous analyses, we now  have access to a complete cluster catalogue covering the vast majority of the M\,83 galaxy. We have estimated the cluster formation efficiency across the whole disc of M\,83 using several techniques and SFR tracers. We have sliced the cluster catalogue in 4 bins either containing the same number of clusters (as done by SV13) or having the same area. Bins containing the same number of clusters remove the size-of-sample effect. However, we show in Section~\ref{sec:results} that our results are not affected by the binning technique.  

In general, we find that, within a factor of two, the SFR derived using \ha\, as tracer or stellar count techniques are in very good agreement. Therefore, we conclude that the M\,83 star formation rate has been nearly constant in at least the last several tens of Myr as suggested in B12, SV13, and SV14. 

We derive lower limits to the cluster formation efficiency, \ga\, as function of galactocentric distance and using the position of the 7 pointings across the galactic body. The CFE of the whole cluster population of M\,83 has a lower limit of $\sim 18$ \%. With a wider coverage of the galaxy and thus a more numerous cluster population we are able to put strong constraints on the derived \ga\, values. We see a net decline in \ga\, from about 26\% in the inner bin to 8\% in the outer bin. A similar decline is observed in the averaged gas surface density and in the predictions made with the fiducial model by \citet{2012MNRAS.426.3008K}. We see that the derived values of \ga\, versus $\Sigma_{\rm SFR}$ in different regions of M\,83 follow the \ga-$\Sigma_{\rm SFR}$ relation of galaxies. We conclude that the relation appears to hold not only on galactic but also on sub-galactic scales. The dependence of $\Gamma$ on $\Sigma_{\rm SFR}$ arises from a more fundamental dependence on the gas pressure (or surface density) through the Schmidt-Kennicutt relation. Therefore, the amount of stars locked in clusters appears to be regulated by the same mechanism that regulates the overall star formation process. Similarly to the Schmidt-Kennicutt relation, a well-calibrated $\Gamma$-$\Sigma_{\rm SFR}$ or $\Gamma$-$\Sigma({\rm H}_2)$ relation can potentially be used to make realistic predictions of cluster formation in cosmological simulations \citep[e.g.][]{2012MNRAS.426.3008K}. However, more effort is needed it to empirically link cluster properties to star formation and GMC properties.

Another aspect that can potentially reveal imprints of the galactic environment on the cluster population is the value of the possible truncation in the upper mass-end of the ICMF. To investigate this issue we have built the ICMFs of the cluster population contained in each galactocentric radius bin. We observe a significant steepening of the ICMF as function of galactocentric distances. Monte Carlo simulations of a single power-law function without upper mass limit are not able to reproduce the observed ICMFs. The steepening can only be consistently reproduced if a Schechter function  of slope $-2$ and exponential cut-off at the high-mass end with a varying \mc\ is used. \mc\, decreases significantly in the outer radial bins. This finding is consistent with the observed decrease of the $\Sigma({\rm H}_2)$ and likely of the maximum GMC mass. The probability that the ICMF of M\,83 is drawn by a single power-law mass function of slope $-2$ is $\sim 4\times10^{-3}$. The probability increases significantly if a steeper power-law function ($\alpha\approx-2.2$) is used. The best representation of the M\,83 ICMF is a Schechter function of slope $-2$ and \mc$\approx1.5\times10^5$ \msun. Overall we conclude that the young star cluster mass function can be described by an almost universal ICMF of slope $-2$ and a variable \mc\, which appears to be a function of the host galaxy environment. 
Upcoming systematic surveys of stellar cluster populations will enable us to quantify and constrain this variation further \citep[e.g.,][]{2015AJ....149...51C}. 
 
\section*{Acknowledgments}
We are deeply indebted to Cliff Johnson for his contribution to the latest stages of the draft. We thank A. Lundgren for providing the fully reduced CO map of M\,83 and and the anonymous referee for providing useful suggestions which improved the draft. AA is thankful to prof. Th. Henning and the Max Planck institute for Astronomy in Heidelberg for early support on this project. NB is partially funded by a Royal Society University Research Fellowship. ESV acknowledge Estrategia de Sostenibilidad 2014-2015 de la Universidad de Antioquia. JER gratefully acknowledges the support of the National Space Grant College and Fellowship Program and the Wisconsin Space Grant Consortium. The results presented here are partially based on observations made with the NASA/ESA Hubble Space Telescope, and obtained from the Hubble Legacy Archive, which is a collaboration between the Space Telescope Science Institute (STScI/NASA), the Space Telescope European Coordinating Facility (ST-ECF/ESA) and the Canadian Astronomy Data Centre (CADC/NRC/CSA).

\appendix
\section{The CMF of class1 \& 2 clusters in M83}
We report here the CMF analysis of class1 and 2 objects contained in bins of equal area.
\begin{figure*}
\centering
\includegraphics[width=5cm]{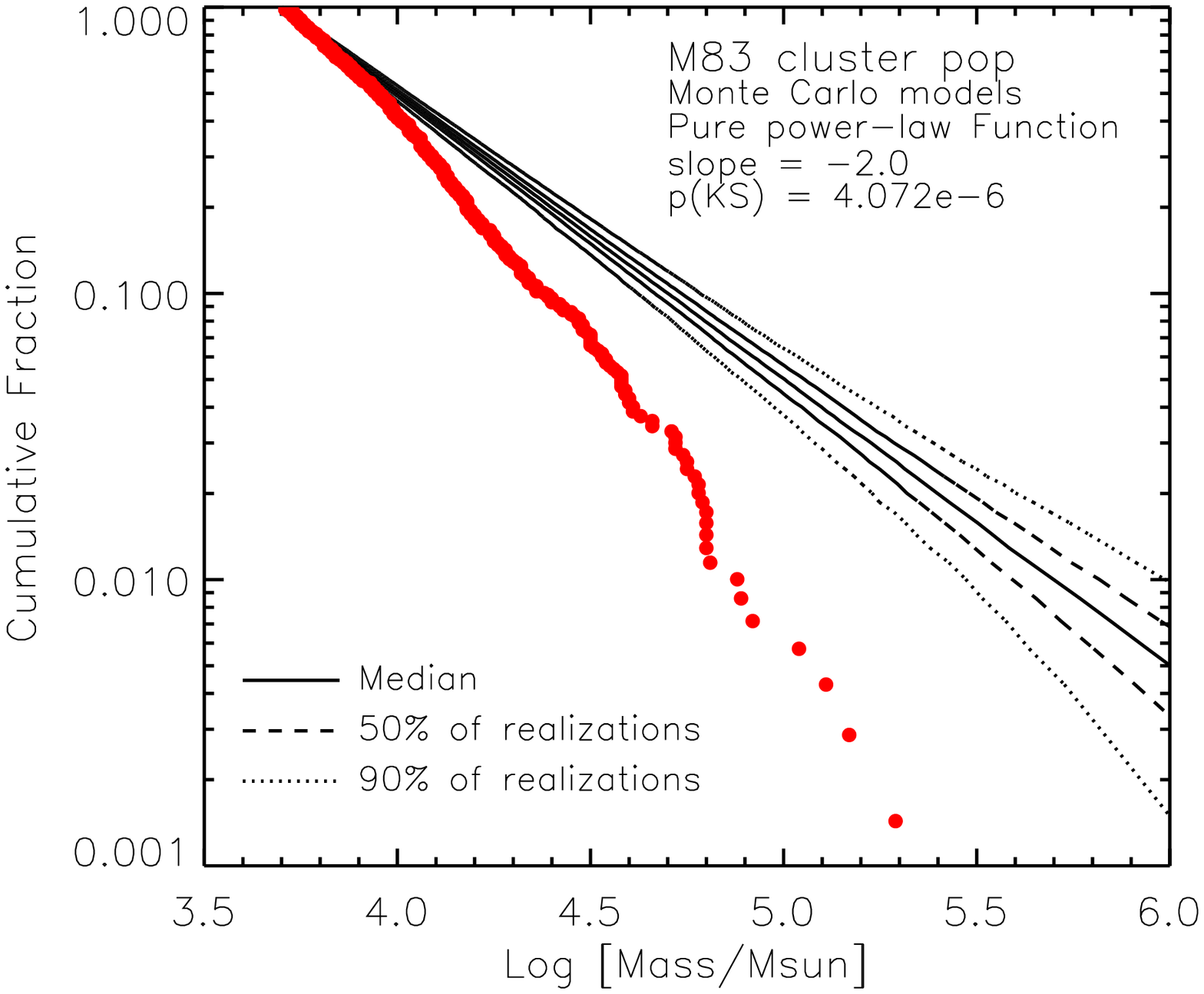}
\includegraphics[width=5cm]{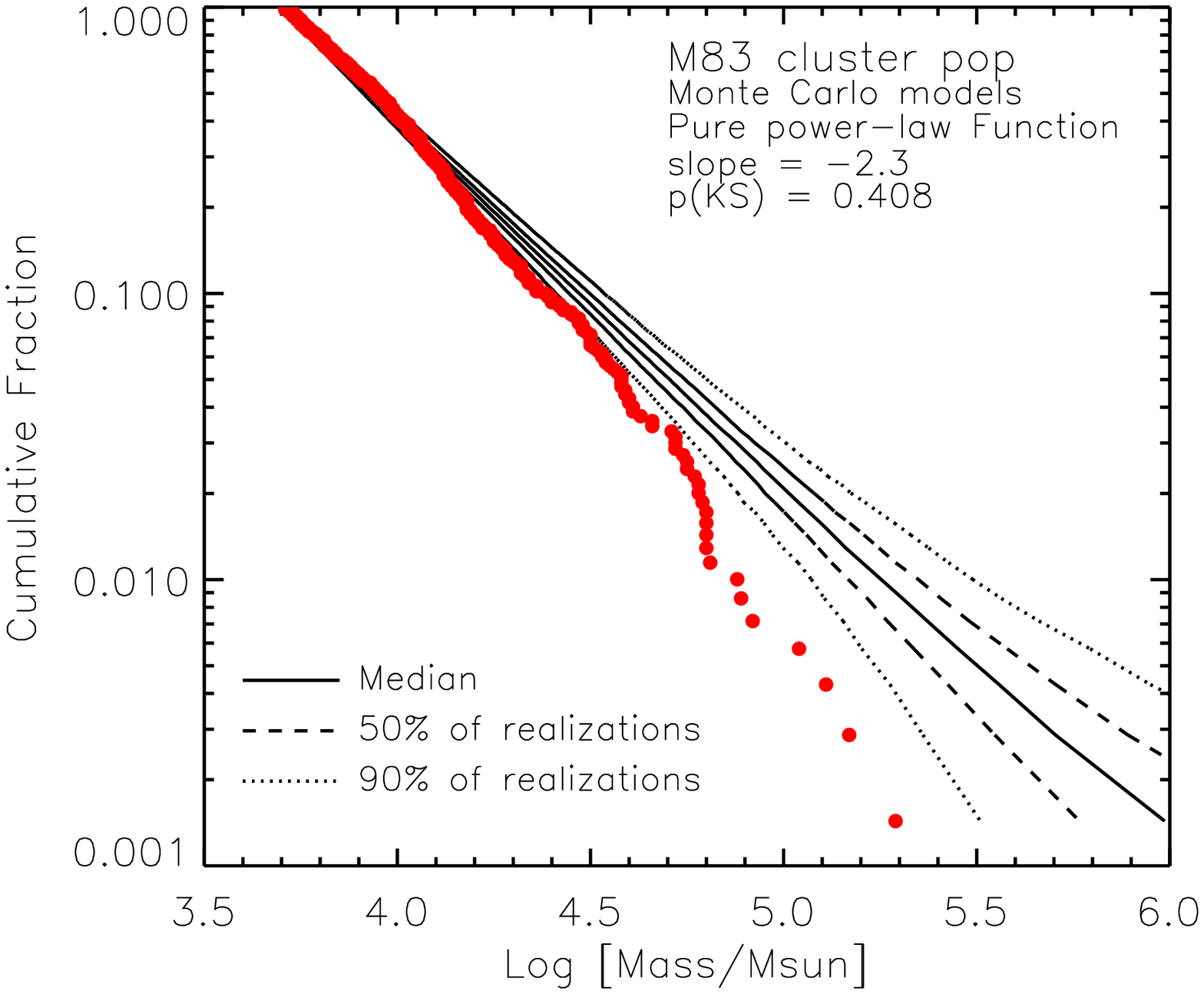}
\includegraphics[width=5cm]{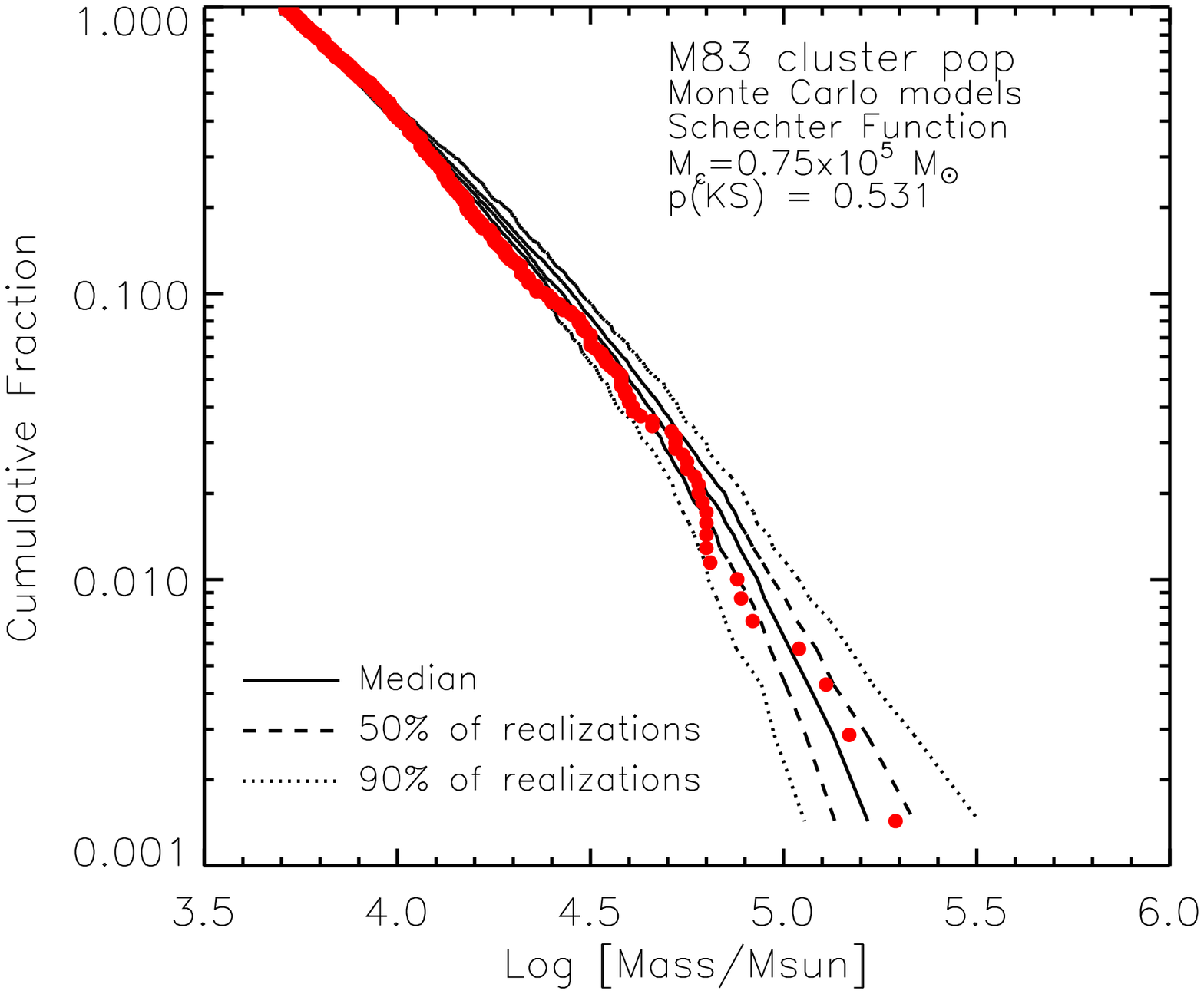}\\
\includegraphics[width=5cm]{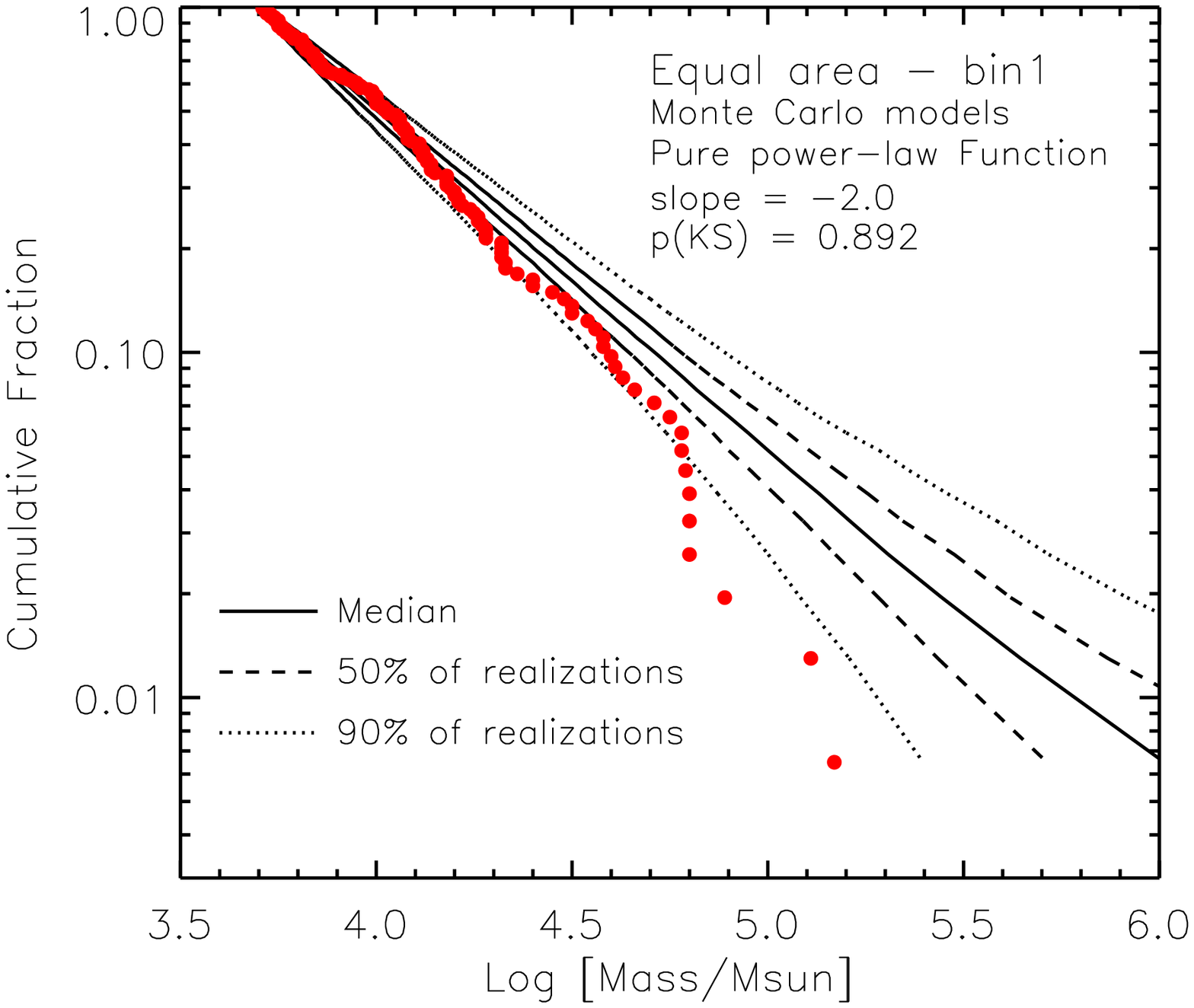}
\includegraphics[width=5cm]{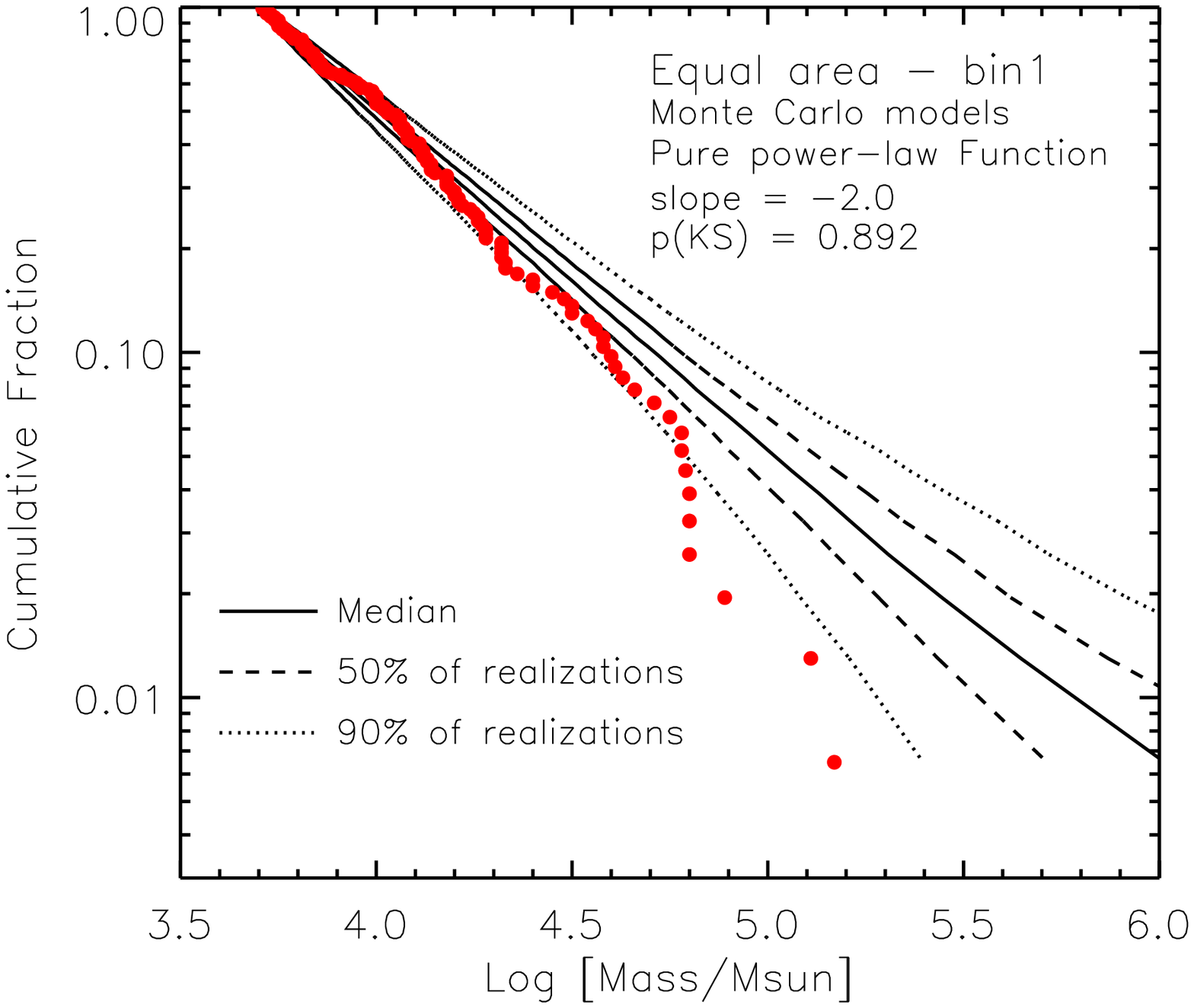}
\includegraphics[width=5cm]{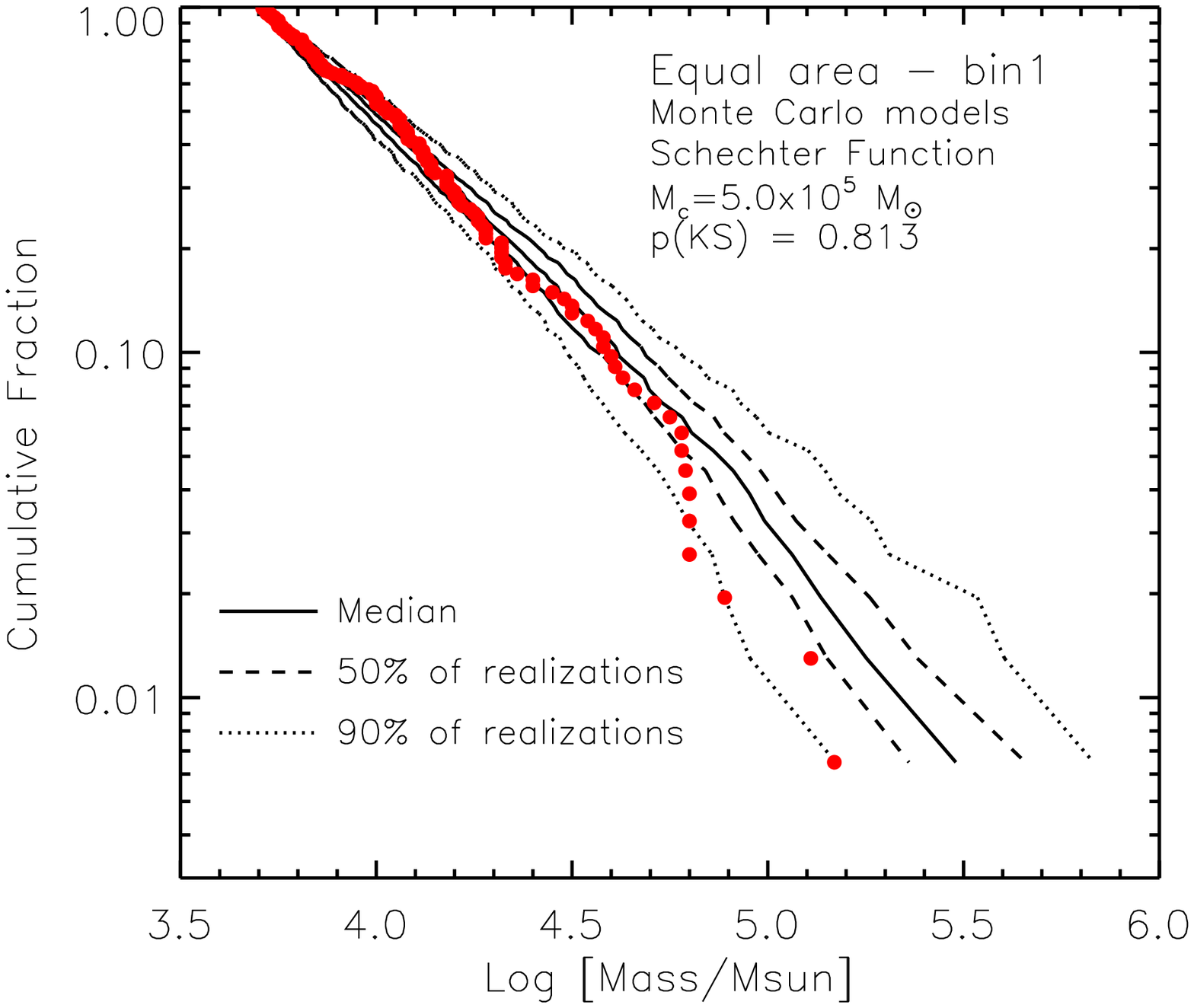}\\
\includegraphics[width=5cm]{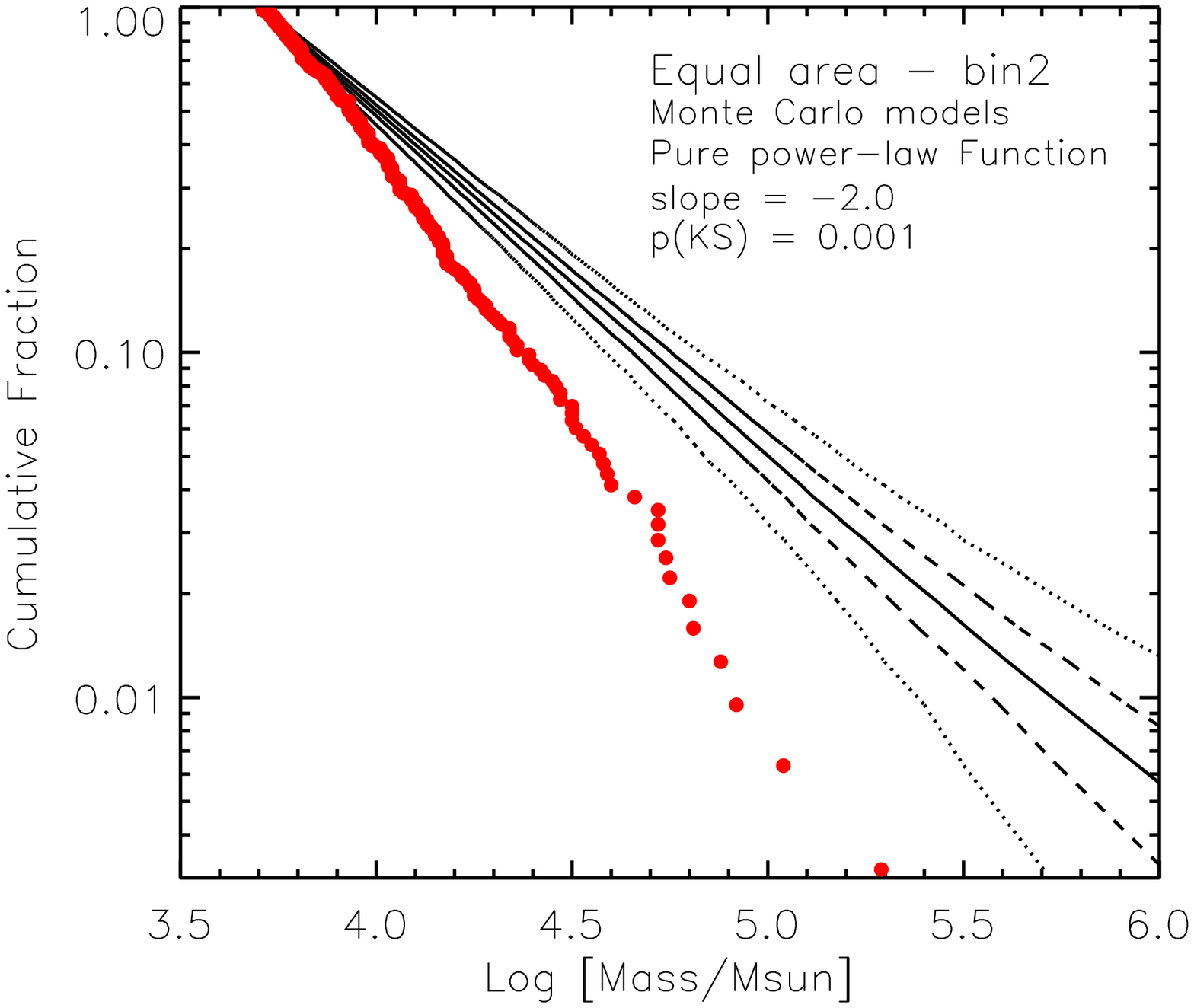}
\includegraphics[width=5cm]{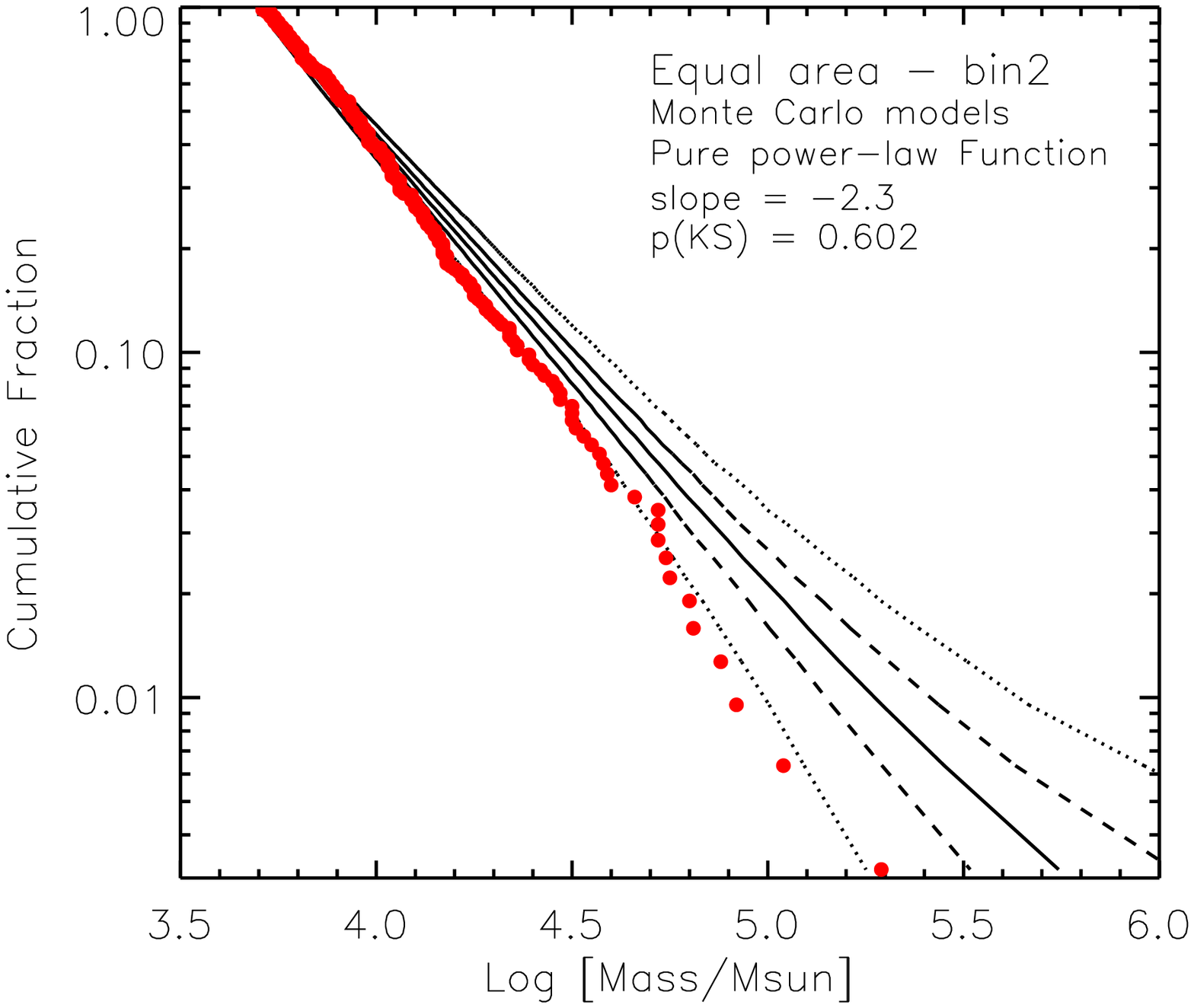}
\includegraphics[width=5cm]{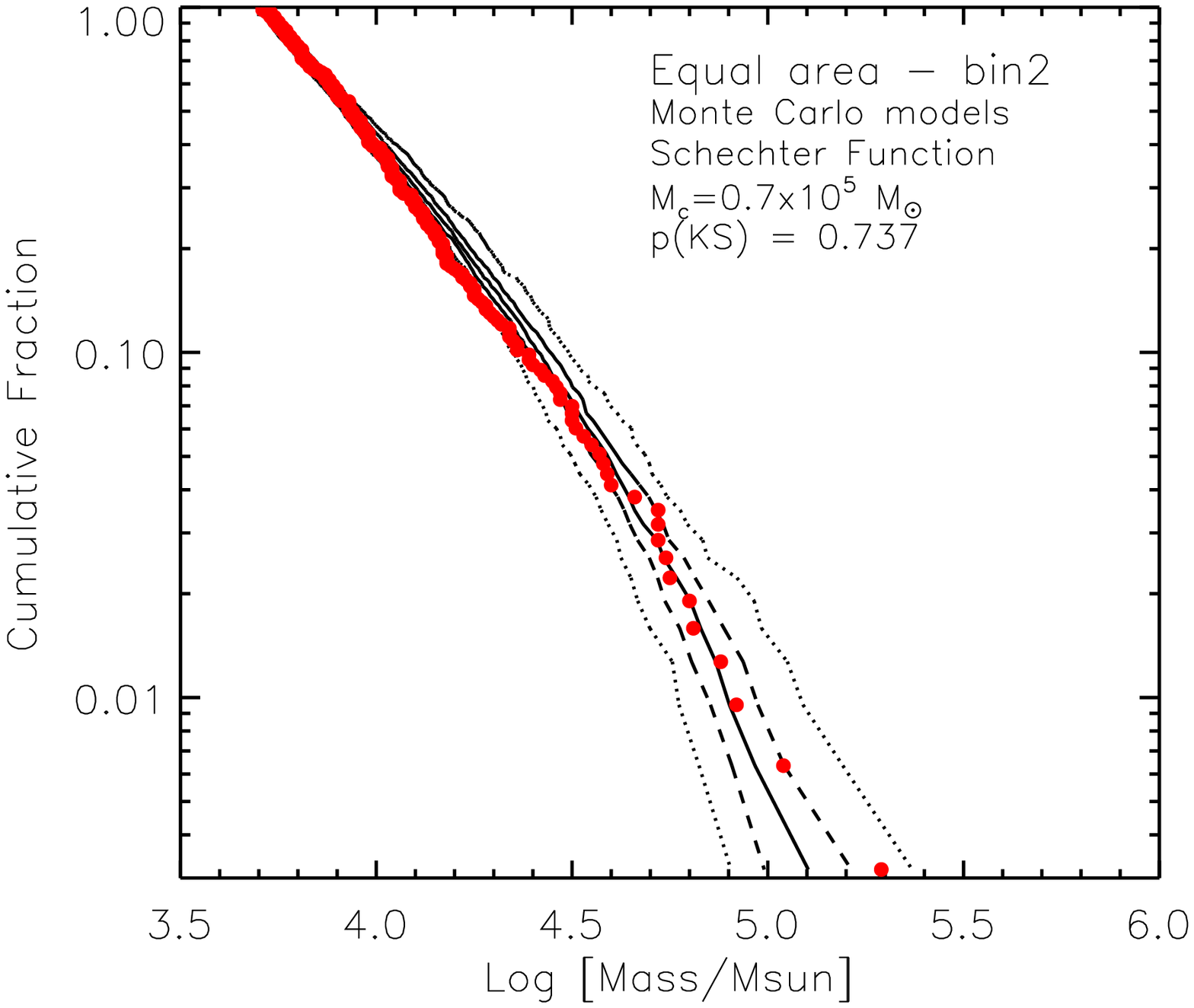}\\
\includegraphics[width=5cm]{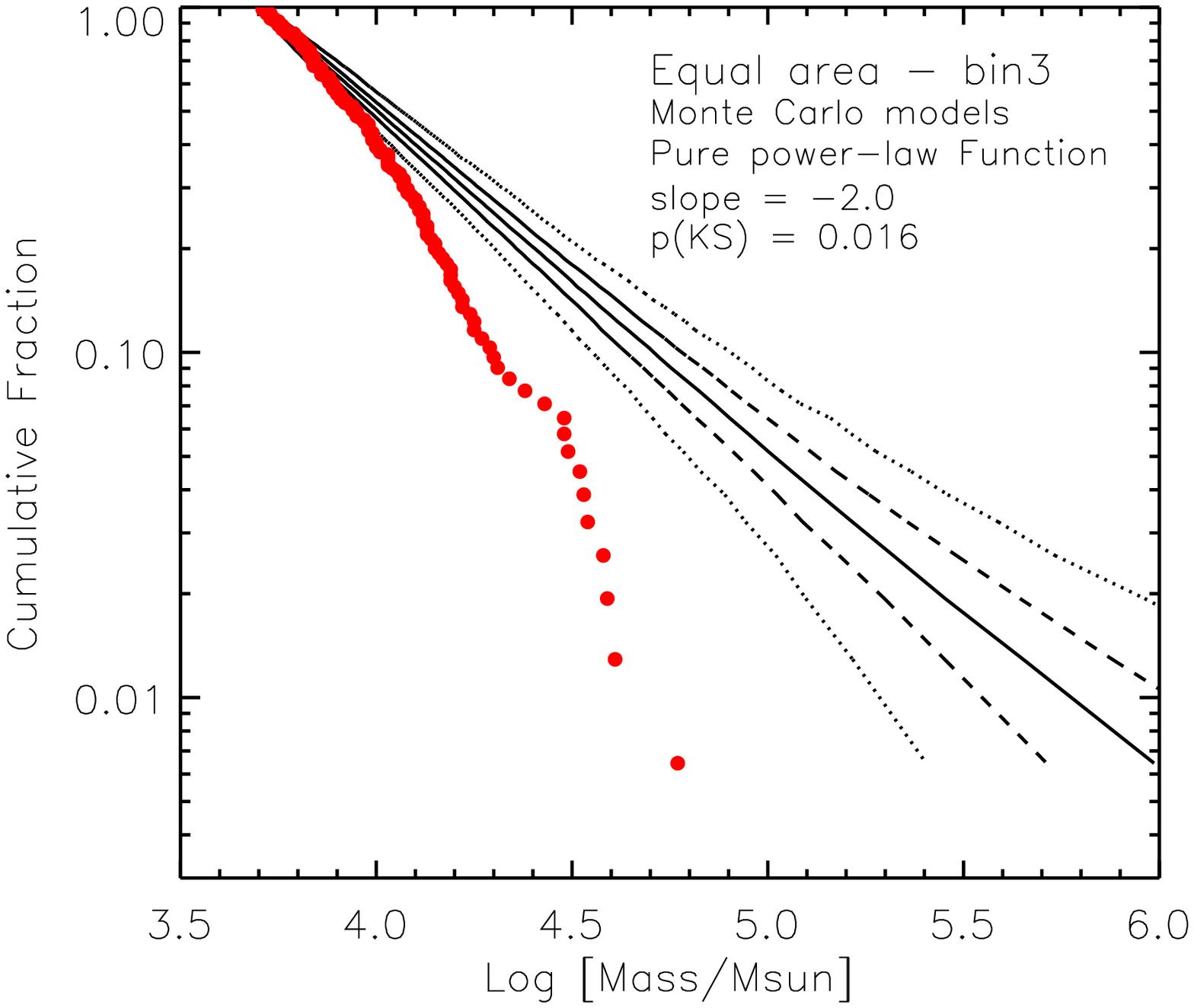}
\includegraphics[width=5cm]{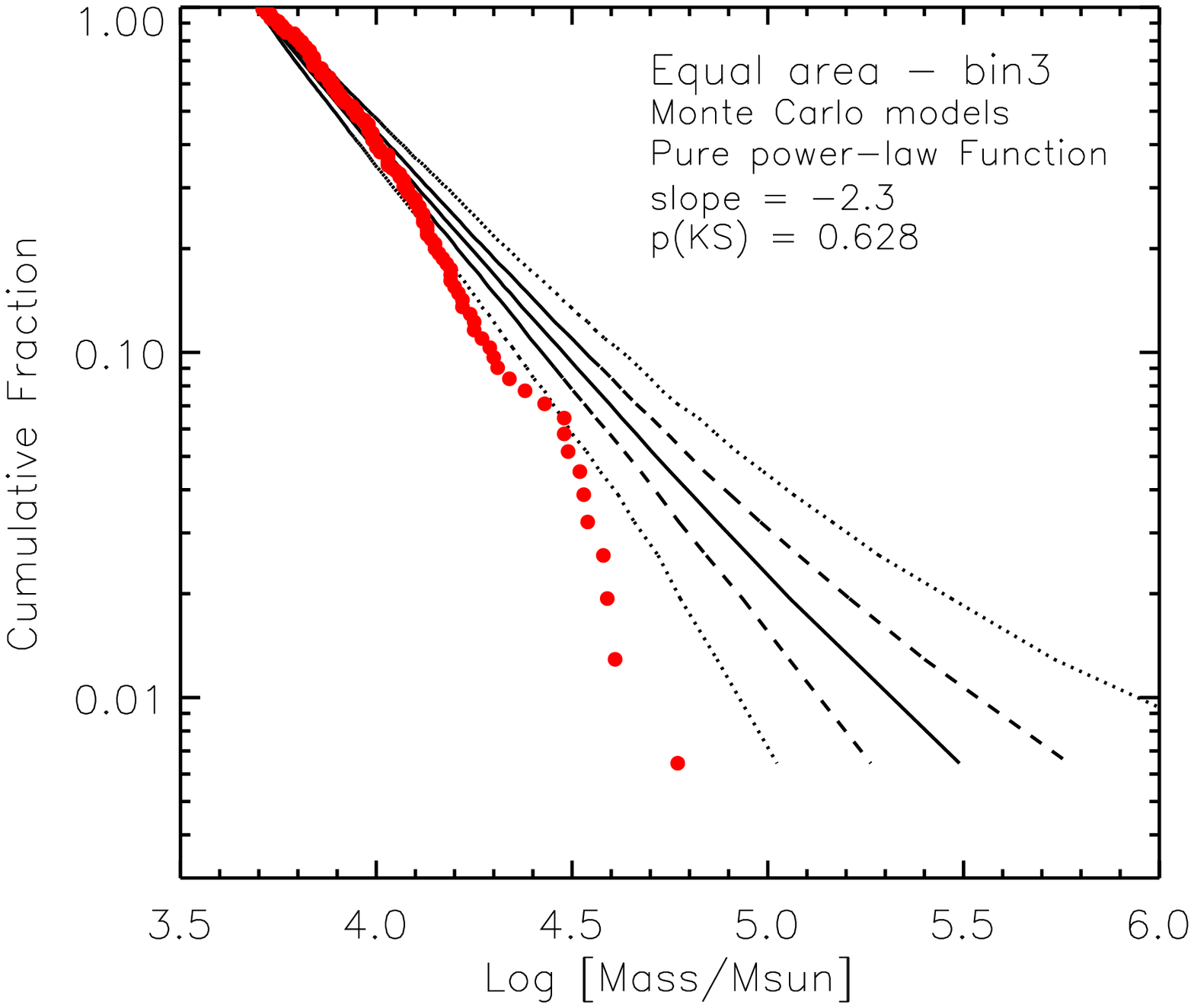}
\includegraphics[width=5cm]{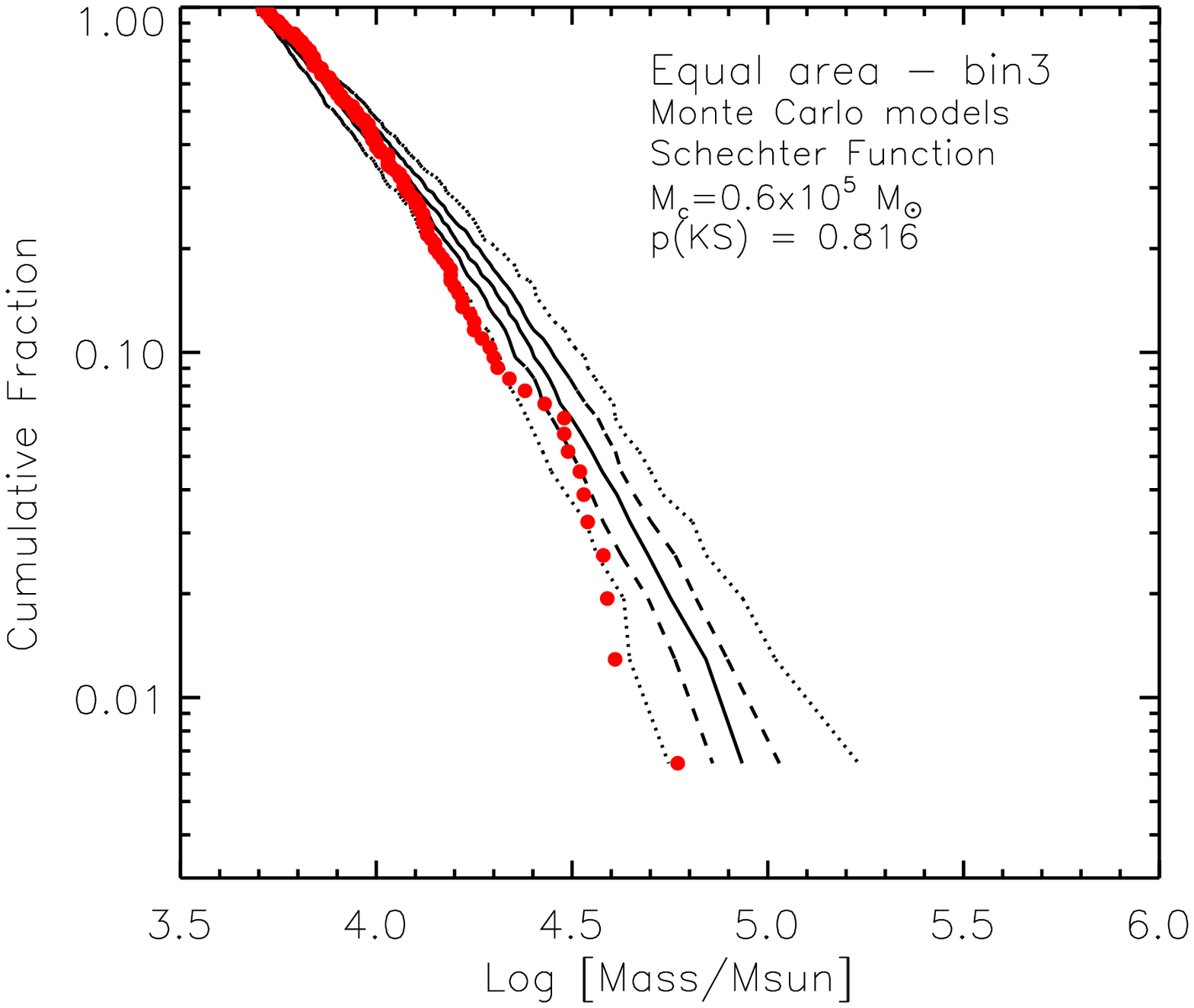}\\
\includegraphics[width=5cm]{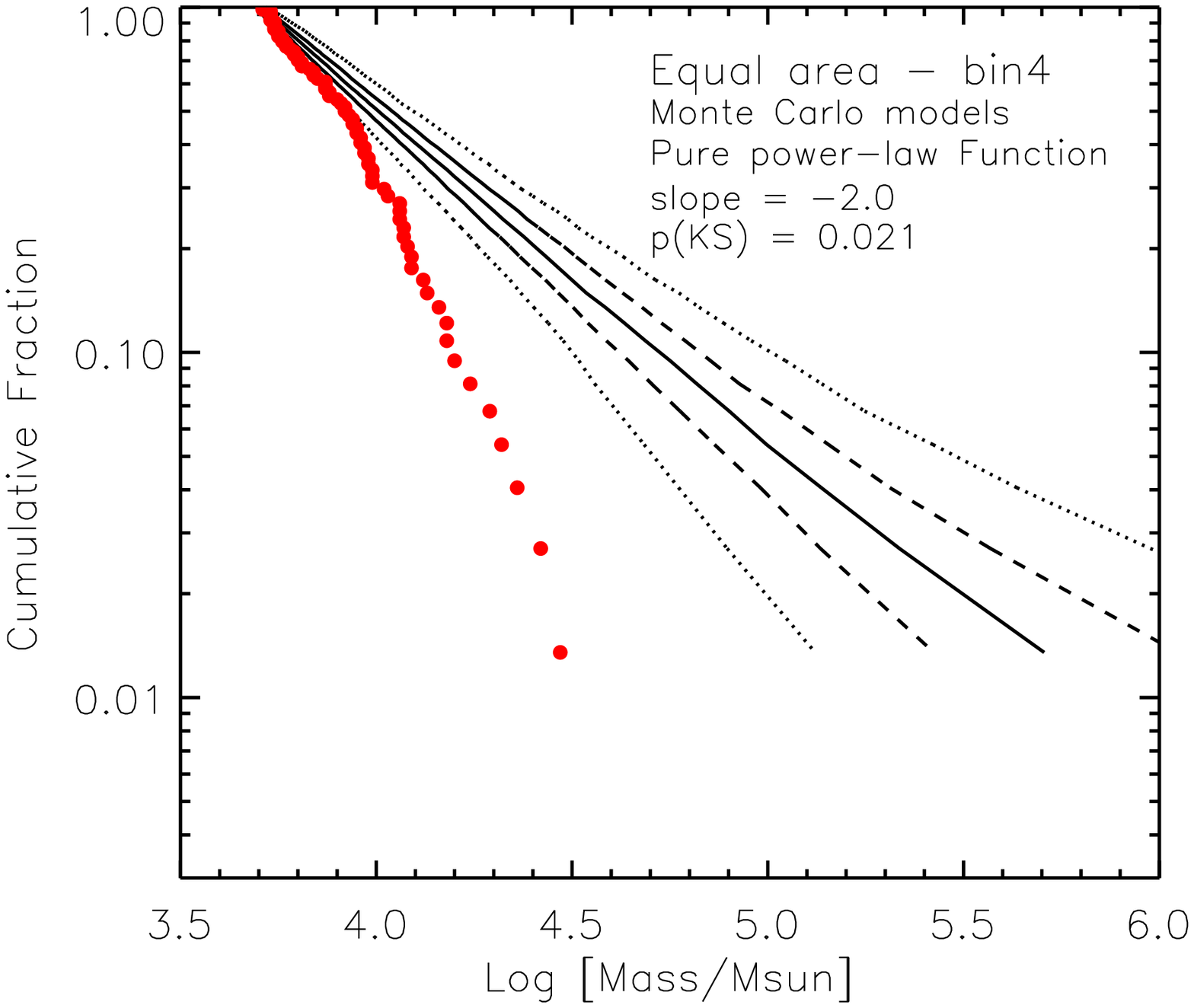}
\includegraphics[width=5cm]{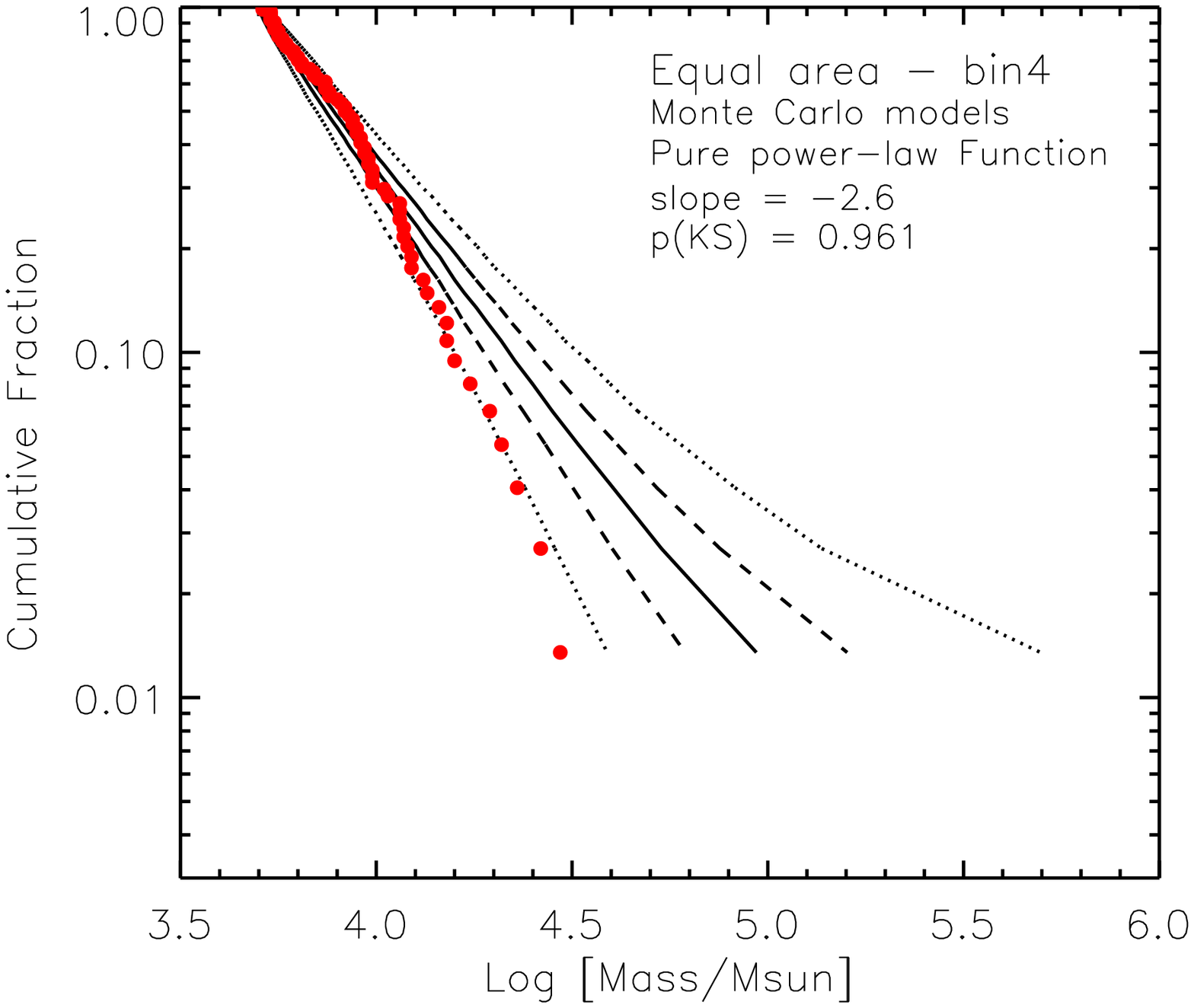}
\includegraphics[width=5cm]{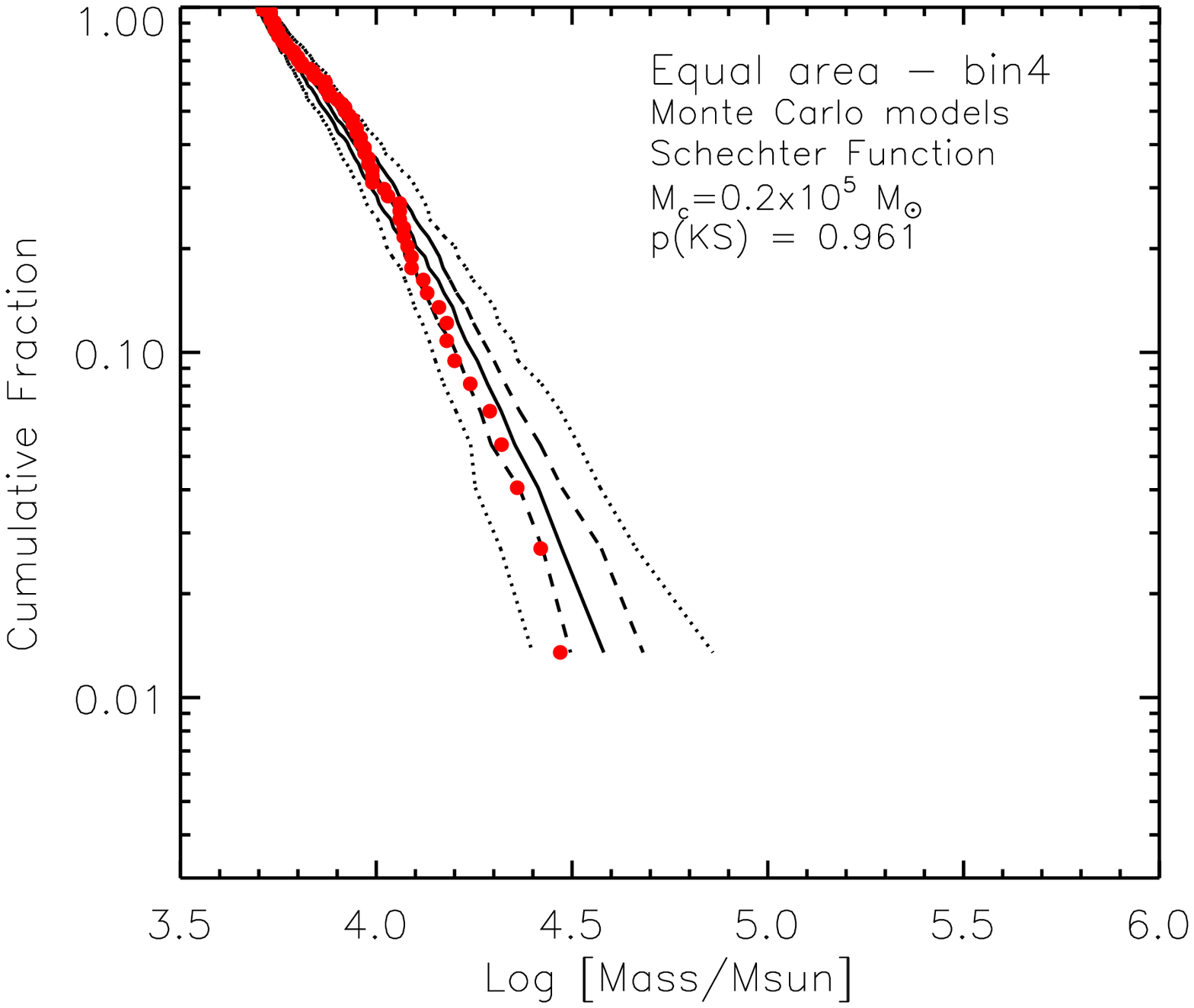}\\
\caption{Cumulative ICMFs (red filled dots) of the whole population and of the clusters within bins of same area (from top to bottom). Both class1 and class 2 systems more massive than 5000 \msun\, and ages between 3 and 100 Myr have been included. Median (solid line), quartiles (dashed line) and extended boundaries (dotted line) of Monte Carlo simulations of cluster populations with the same number of objects as the observed ones are included in each panel. In the first column, to generate the fake populations we assume a pure power-law ICMF with slope $-2$. In the middle column we still assume a power-law ICMF but with varying slopes (see insets), chosen to maximise the Kolmogov-Smirnov probability that the observed and simulated distributions are drawn from the same parent distribution, p(KS). In the right column we assume that the ICMF is better described by a Schechter function of slope $-2$ and a exponential cut-off, \mc, chosen to maximise the KS probability, as indicated in the inset. }
\label{fig:mf_rad_app}
\end{figure*} 
\onecolumn
\section{A Complete compilation of the data samples available in the literature}
Complete compilation of the data available in the literature and plotted in Figure~\ref{fig:mv_sfr} and \ref{fig:gamma_sfr}. The footnote associated with the name of the galaxy gives the reference for all the values reported on the corresponding row, unless a different note on the values specify a different source. Some galaxies are reported more than once, with different values of SFR and $\Sigma_{\rm SFR}$, because \ga\, and related quantities (SFR, and $\Sigma_{\rm SFR}$) have been derived using only partial coverages of galaxies.

\vspace{0.5cm}
\begin{table}
 \begin{center}
       \caption{$a$) \citet{2011MNRAS.417.1904A} (blue star symbols in Figure~\ref{fig:mv_sfr}); $b$) \citet{2002AJ....124.1393L} (black triangles in Figure~\ref{fig:mv_sfr}), only a fraction of the original catalogue is listed while the values of the omitted ones are listed with the most recent estimated values available in the literature; $c$) The values used for each binning (purple filled dots in Figure~\ref{fig:mv_sfr}) are reported in Table~\ref{tab2} of the main text; $d$) \citet{2008MNRAS.390..759B} (black squares in Figure~\ref{fig:mv_sfr}); $e$) \citet{2014AJ....147...78W} (orange diamonds in Figure~\ref{fig:mv_sfr}), some of the galaxies in the original list have been omitted because \citet{2002AJ....124.1393L} presented data with a larger coverage of the galaxy than in this paper. The dwarf galaxies (green star symbols in Figure\ref{fig:mv_sfr}) include datapoints from the following papers: $f$) \citet{2011AJ....142..129A}; $g$) \citet{2011AJ....141..132P}; $h$) \citet{2010MNRAS.405..857G}; $i$) \citet{2009AJ....138..169A}; $j$) \citet{2005AJ....129.2701R}; $k$) \citet{2000AJ....120.1273J}; $l$) \citet{2012ApJ...751..100C}; $m$) \citet{2015ApJ...804..123L}. $o$) In Figure~\ref{fig:gamma_sfr}, we show the combined \ga\, derived by \citet{2012ApJ...751..100C} using all the clusters younger than 10 Myr detected in their dwarf sample. The SFR density reported is not a single value but a range enclosing all the single galactic values. $p$) \citet{2011A&A...529A..25S}; $q$) \citet{2014AJ....148...33R}; $r$) This work. Values obtained considering the age range 1-10 Myr, see Table~2 in the main text for more details.}
\label{tab5}
  \begin{tabular}{|c|c|c|c|c|}
  
\hline
Galaxy & M$_{V}^{\rm bright}$& SFR & $\Sigma_{\rm SFR}$ & \ga \\
 	& [mag] & M$_\odot$yr$^{-1}$&M$_\odot$yr$^{-1}$Kpc$^{-2}$& \% \\
\hline
ESO\,338$^{a}$  &   -15.50  &    3.20     & 1.55 & 50.0$\pm 10.0$\\     
Haro\,11$^{a}$   &  -16.16   &  22.00    &  2.16& 50.0$^{+13.0}_{-15.0}$\\
ESO\,185-IG13$^{a}$ &    -14.36  &   6.40  &   0.52& 26.0$\pm 5.0$\\
MRK\,930$^{a}$   &  -15.00   &   5.34   &  0.59 & 25.0$\pm 10.0$\\
SBS\,0335-052E$^{a}$  &   -14.28   &   1.30  &   0.95 & 49.0$\pm 15.0 $\\
NGC\,247$^{b}$  &       -10.20&       0.0360  &     0.18$\times10^{-3}$  &  --   \\
NGC\,300$^{b}$  &       -9.90 &       0.0779  &     0.49$\times10^{-3}$  &  --   \\
NGC\,1156$^{b}$ &       -11.10 &        0.1842  &      3.07$\times10^{-3}$  &  --   \\
NGC\,2403$^{b}$  &       -9.90  &        0.3376  &     0.97$\times10^{-3}$  &  --   \\
NGC\,2835$^{b}$  &       -10.90  &       0.0920  &     0.73$\times10^{-3}$  &  --   \\
NGC\,2997$^{b}$  &       -12.90  &         1.8604  &      3.07$\times10^{-3}$  &  --   \\
NGC\,2997$^q$  &   -- & -- & 9.4$\times10^{-3}$ & 10.0$\pm$2.6 \\
NGC\,3184$^{b}$  &       -10.60  &        0.3956  &      1.72$\times10^{-3}$  &   --  \\
NGC\,3621$^{b}$  &       -11.90  &        0.8801  &      1.67$\times10^{-3}$  &  --   \\
NGC\,5204$^{b}$ &       -9.60  &       0.0315  &     0.83$\times10^{-3}$  &  --   \\
M\,83$^{b, c}$  &       -11.70  &         2.2842  &       13.76$\times10^{-3}$  &  --   \\
M\,83 (centre)$^h$ &  -- & 0.39  & 0.54 & 26.7$^{+5.3}_{-4.0}$ \\
M\,83 (outer)$^p$   &  --  & 0.39  & 0.013 &  5.6$\pm0.6$\\
M\,83 (0.45-4.5 kpc)$^r$ & -11.60 & 0.82 & 0.013 & 18.2$\pm$3.0 \\
NGC\,5585$^{b}$  &       -10.80  &       0.0336  &     0.32$\times10^{-3}$ &  --   \\
NGC\,6744$^{b}$  &       -11.00  &        0.4309  &     0.62$\times10^{-3}$  &  --   \\
NGC\,6946$^{b}$  &       -13.00  &         2.5392  &      4.60$\times10^{-3}$  &  --   \\
NGC\,6946$^{h}$(WFPC2)  &       --  &         0.17  &      4.60$\times10^{-3}$  &   12.5$^{+1.8}_{-2.5}$ \\
NGC\,7424$^{b}$  &       -11.40  &        0.1728  &     0.18$\times10^{-3}$  &  --   \\
NGC\,1741$^{b}$ &       -15.00  &         4.8819  &       12.78$\times10^{-3}$  & --    \\
NGC\,5253$^{b}$  &       -11.10  &        0.2114  &      7.29$\times10^{-3}$  &  --   \\
IC\,1613$^{b}$  &       -5.80  &     0.0004  &     0.05$\times10^{-3}$  &  --   \\
LMC &       -11.1$^{h}$   &        0.1201$^{b}$   &      1.52$\times10^{-3}$$^{b}$   &  5.8$\pm0.5$$^{h}$  \\
NGC\,4214$^{b}$  &       -12.04  &       0.0798  &      3.80$\times10^{-3}$  &  --   \\
DDO\,50$^{b}$  &       -7.91  &       0.0108  &      1.26$\times10^{-3}$  &  --   \\
DDO\,168$^{b}$  &       -7.58  &      0.0043  &     0.85$\times10^{-3}$  &  --   \\
DDO\,165$^{b}$  &       -8.34  &     0.0005  &     0.18$\times10^{-3}$  &  --   \\
Sextans A$^{b}$  &       -7.12  &      0.0037  &      2.29$\times10^{-3}$  &  --   \\
NGC\,3521$^{b}$  &       -11.50  &         1.3676  &      3.58$\times10^{-3}$  &  --   \\
NGC\,4258$^{b}$  &       -12.60  &        0.9926  &     0.70$\times10^{-3}$  &  --   \\
NGC\,5055$^{b}$  &       -11.40  &         1.5019  &      2.98$\times10^{-3}$  &  --   \\
M\,51$^{b}$  &       -12.80  &         4.7454  &      8.21$\times10^{-3}$  &  --   \\
NGC\,7252$^{d}$   &       -13.40  &        5.40  &   --  &  --  \\
NGC\,6240$^{d}$   &       -16.40  &        140.00  &   --  & --   \\
NGC\,2207$^{d}$   &       -13.60  &        2.20  &   --  &  --  \\
NGC\,1275$^{d}$   &       -15.30  &        12.40  &   --  &  --  \\
  \hline
\end{tabular}

\end{center}
\end{table}
\begin{table}
 \begin{center}
  \begin{tabular}{|c|c|c|c|c|}

\multicolumn{5}{c}{{\bf Table B1.}  -- Continued}\\
\multicolumn{5}{c}{    }\\
\hline
Galaxy & M$_{V}^{\rm bright}$& SFR & $\Sigma_{\rm SFR}$ & \ga \\
 	& [mag] & M$_\odot$yr$^{-1}$&M$_\odot$yr$^{-1}$Kpc$^{-2}$& \% \\
\hline
M\,82$^{d}$   &       -14.80  &        7.00 &   --  &  --  \\
NGC\,3597$^{d}$   &       -13.30  &        10.80  &   --  & --   \\
IRAS\,19115$^{d}$   &       -16.80  &        192.00  &   --  &  --  \\
NGC\,1533$^{d}$ (A1)   &       -7.17  &    0.37$\times10^{-3}$  &   --  & --   \\
NGC\,1533$^{d}$ (A2)  &       -5.71 &    0.25$\times10^{-3}$  &   --  &  --  \\
NGC\,1533$^{d}$ (A3)   &       -6.16  &    0.18$\times10^{-3}$ &   --  & --   \\
NGC\,2623$^{d}$   &       -14.50  &        51.00  &   --  &  --  \\
NGC\,3256$^{d}$   &       -15.70  &        46.00  &   0.62  & 22.9$^{+7.3}_{-9.8}$$^h$  \\
NGC\,7673$^{d}$   &       -14.70  &        4.90  &   --  & --   \\
NGC\,6745$^{d}$   &       -15.00  &        12.20  &   --  &  --  \\
NGC\,1140$^{d}$   &       -14.80  &       0.80  &   --  & --   \\
Milky Way$^h$ & -- & 0.1508  & 0.012 & 7.0$^{+7.0}_{-3.0}$\\
NGC\,45$^{e}$  & -10.83  &   0.12  &  --  & --  \\
NGC\,45$^{p}$  & -- &       0.05  &  1.02$\times10^{-3}$  & 5.2$\pm$0.3 \\
NGC\,406$^{e}$  &       -11.75  &       0.29  &  --  & --  \\
NGC\,628$^{e}$  &       -11.84  &       0.23  &  --  & --  \\
NGC\,1300$^{e}$ (F1)  &       -11.00  &       0.30  &  --  &  -- \\
NGC\,1300$^{e}$ (F2)  &       -11.52  &       0.32 &  --  &  -- \\
NGC\,1309$^{e}$  &       -13.80  &        1.70  &  --  &  -- \\
NGC\,1313$^{e}$  &       -10.98  &       0.22  &  --  &  -- \\
NGC\,1313$^{p}$  &       --  &   0.68  &  0.011  & 3.2$\pm$0.2 \\
NGC\,1483$^{e}$  &       -10.01  &       0.11  &  --  & --  \\
NGC\,3627$^{e}$  &       -11.97  &       0.37  &  --  &  -- \\
NGC\,4038$^{e}$  &       -15.25  &        2.43  &  --  & --  \\
NGC\,4394$^{e}$  &       -10.25  &       0.14  &  --  &  -- \\
NGC\,4395$^{e}$  &       -9.79  &      0.07  &  --  &  -- \\
NGC\,4395$^{p}$  &       --  &      0.17  &  4.66$\times10^{-3}$  &  1.0$\pm$0.6\\
NGC\,4736$^{e}$  &       -10.44  &      0.04  &  --  &  -- \\
NGC\,5055$^{e}$  &       -9.61  &      0.02  &  --  & --  \\
M\,101$^{e}$ (F1)  &       -11.38  &       0.25  &  --  & --  \\
M\,101$^{e}$ (F2)  &       -11.57  &       0.20  &  --  & --  \\
NGC\,6503$^{e}$&       -10.51  &      0.07  &  --  & --  \\
NGC\,7793$^{e}$  &       -9.65  &      0.07  &  --  & --  \\
NGC\,7793$^{p}$  &     -- &      0.15  &  6.51$\times10^{-3}$   & 2.5$\pm$0.3  \\
NGC\,4449$^{f}$           & -12.60 &   1.0	&   0.04 &  9.0 \\ 
NGC\,1569    &   -13.9$^{b}$  & 0.36$^{g}$  &  0.03$^{g}$  &  13.9$\pm0.8^h$ \\ 
NGC\,1705$^{i}$  &   -13.8  & 0.31 & 0.046  & --   \\
SMC$^j$  & -9.94 &  0.043   &  0.001    & 4.2$^{+0.2}_{-0.3}$$^h$  \\ 
He2-10$^k$  & -12.5  &  0.20 &  0.20 &  --  \\  
NGC\,2366$^l$ &-8.52  &  0.094 &  2.4$\times10^{-3}$ &--  \\       
UGC\,4305$^l$& -8.88  & 0.114  &  2.2$\times10^{-3}$   & -- \\ 
UGC\,4459$^l$ &  -7.9   &    0.003 &  1.0$\times10^{-3}$ & -- \\   
UGC\,5336$^l$ &  -8.59  &   0.004 & 0.00037 & -- \\  
IC\,2574$^l$ & -9.12  &   0.11  &  1.2$\times10^{-3}$  & -- \\      
UGC\,5692$^l$ &-9.44  &   7.34$\times10^{-3}$ & 0.5$\times10^{-3}$   &--  \\  
UGCa\,292$^l$  &  -7.46  &   5.15$\times10^{-3}$ & 0.01   &--  \\     
UGC\,8201$^l$  &  -9.16  &   31.6$\times10^{-3}$  & 2.1$\times10^{-3}$  --   & \\  
UGC\,9128$^l$  & -5.45  &   0.3$\times10^{-3}$ & 0.3$\times10^{-3}$ &--  \\   
UGC\,9240$^l$  &  -5.90  &  5.24$\times10^{-3}$ & 1.0$\times10^{-3}$  &--  \\ 
Dwarf sample$^o$ & -- & -- & [0.01, 10.0]$\times10^{-3}$ & 5.0 \\
IC\,10$^m$       &  -10.4	&  0.07  &	0.03   & 4.2 \\      

  \hline
\end{tabular}

\end{center}
\end{table}
\twocolumn
\label{lastpage}
\end{document}